\relax
\documentclass[article]{elsarticle}
\usepackage[margin=1in]{geometry}
\usepackage{lineno,hyperref}
\modulolinenumbers[5]
\usepackage{times}

\usepackage{soul}
\usepackage{url}
\usepackage[utf8]{inputenc}
\usepackage[small]{caption}
\usepackage{graphicx}
\usepackage{xcolor}
\usepackage{booktabs}
\usepackage{amsthm, amsmath, amssymb}
\usepackage{mathtools}
\usepackage{graphicx}
\usepackage{subcaption,float}
\usepackage{breqn}
\usepackage[autostyle]{csquotes}
\usepackage[british]{babel}
\usepackage{breakcites}

\journal{Econometrics and Statistics}

\usepackage{booktabs}
\usepackage{amssymb}
\usepackage{amsmath}
\usepackage{bm}
\usepackage{enumerate}
\usepackage{algorithm}
\usepackage{algorithmic}
\usepackage{multirow}
\usepackage{verbatim}
\usepackage{caption}
\usepackage{subcaption}
\usepackage{tabularx}
\usepackage{listings}

\def\bx{\mathbf{x}}
\def\by{\mathbf{y}}
\def\bz{\mathbf{z}}

\def\bu{\mathbf{u}}

\def\bF{\mathbf{F}}

\def\bX{\mathbf{X}}

\def\bU{\mathbf{U}}
\def\bV{\mathbf{V}}

\def\b0{\mathbf{0}}
\def\b1{\mathbf{1}}

\def\bmu{\mbox{\boldmath $\mu$}}

\def\btheta{\mbox{\boldmath $\theta$}}

\def\bSigma{\mbox{\boldmath $\Sigma$}}

\def\bPsi{\mbox{\boldmath $\Psi$}}
\def\bTheta{\mbox{\boldmath $\Theta$}}

\DeclareMathOperator*{\argmax}{arg\,max}

\newcommand{\vecx}{\mathbf{x}}

\newcommand{\vecy}{\mathbf{y}}

\newcommand{\vecmu}{\mbox{\boldmath$\mu$}}

\newtheorem{lemma}{Lemma}

\biboptions{authoryear}

\begin{document}

\title{
Improved Inference of Gaussian Mixture Copula Model for Clustering and Reproducibility Analysis using 
Automatic Differentiation
}

\begin{abstract}
Copulas provide a modular parameterization of multivariate distributions that decouples the modeling of marginals from the dependencies between them.
Gaussian Mixture Copula Model (GMCM) is a highly flexible copula that can model many kinds of multi-modal dependencies, as well as asymmetric and tail dependencies.
They have been effectively used in clustering non-Gaussian data and in Reproducibility Analysis, a meta-analysis method designed to verify the reliability and consistency of multiple high-throughput experiments.
Parameter estimation for GMCM is challenging due to its intractable likelihood.
The best previous methods have maximized a proxy-likelihood through a Pseudo Expectation Maximization (PEM) algorithm.
They have no guarantees of convergence or convergence to the correct parameters.
In this paper, we use Automatic Differentiation (AD) tools to develop a method, called AD-GMCM, that can maximize the exact GMCM likelihood.
In our simulation studies and experiments with real data, AD-GMCM finds more accurate parameter estimates than PEM and yields better performance in clustering and Reproducibility Analysis. 
We discuss the advantages of an AD-based approach, to address problems related to monotonic increase of likelihood and parameter identifiability in GMCM.  We also analyze, for GMCM, two well-known cases of degeneracy of maximum likelihood in GMM that can lead to spurious clustering solutions. Our analysis shows that, unlike GMM, GMCM is not affected in one of the cases.
\end{abstract}

%\author{Anonymous Authors}
 \author{Siva Rajesh Kasa, Vaibhav Rajan \\
 Department of Information Systems and Analytics\\
 School of Computing, National University of Singapore}

\maketitle

\section{Introduction}

Model-based clustering is a well-established paradigm for clustering multivariate data
\citep{banfield1993model,melnykov2010finite}.
In this approach, data is assumed to be generated by a finite mixture model where each component represents a cluster.
The rigorous probabilistic framework of mixture models facilitates model building and assessment. 
Through the choice of the component distribution, model-based clustering imposes distributional assumptions on the marginals, along each dimension.
These marginal distributions are often assumed to be identical, e.g., in
Gaussian Mixture Models (GMM) that are widely used in a variety of applications \citep{McLa:Peel:fini:2000}.
However, both assumptions of Gaussian distributed components as well as same distributions in all components are often violated in real data.

Copulas provide a  modular parameterization of multivariate distributions that decouples the modeling of marginals from the dependencies between them.
They have been used to model component distributions within mixture models, e.g.,
by \cite{fujimaki2011online,kosmidis2016model}.
When interest lies mainly in discovering  dependencies, copulas provide an elegant model of dependencies with no restrictive assumptions on the marginals.
Such models have been used 
in {\it dependency clustering} that discovers clusters based on their dependency patterns \citep{rey2012copula,rajan2016dependency,tekumalla2017vine}.
For many applications, including clustering, a semi-parametric model works well in practice, where copulas are used to model dependency patterns, assuming no fixed parametric model for the marginals.

The Gaussian Mixture Copula Model (GMCM) combines the modeling strengths of a copula-based approach and Gaussian Mixture Models.
It allows flexible dependency modeling, especially of non-Gaussian data, and can model many kinds of multi-modal dependencies, notably asymmetric and tail dependencies (as illustrated in figure \ref{fig:gmcm_illustration}).
Unlike mixtures that use copulas as component distributions, GMCM is a specific copula family where the (latent) copula density follows a Gaussian Mixture Model (more details are in \S \ref{sec:background}).
This has considerable advantages for copula-based clustering since
clusters can be inferred directly from the dependencies 
obviating the need for marginal parameter estimation.
The advantages of GMCM-based clustering particularly for non-Gaussian data has been shown by
\cite{tewari2011parametric,compstat,rajan2016dependency,bilgrau2016gmcm,kasa2020gaussian}.

\cite{li2011reproducibility} independently developed a specific case of GMCM for Reproducibility Analysis of genomic experiments that has been widely adopted, e.g., in the ENCODE project \citep{encode2012integrated}.
In Reproducibility Analysis, the input data matrix consists of test statistics or p-values from experiments investigating the same null hypothesis for the subjects involved.
In genomics, it is common to have genes as subjects and p-values from different experiments testing the same hypothesis about the genes as the attribute values.
Large values are assumed to be indicative of the alternative hypothesis.
The experiments may differ in the technologies used, sample sizes, features and confounding factors, among others. 
The aim of the analysis is to identify a group of subjects that are commonly significant across experiments.
A special case of GMCM is developed by \cite{li2011reproducibility}
to address this problem through 
a meta-analysis method that can be used to verify the reliability and consistency of multiple high-throughput genomic experiments.

Maximum Likelihood (ML) parameter estimation for GMCM is recognized to be challenging due to the difficulty of evaluating the GMCM likelihood (detailed in \S \ref{sec:paraminf}).
As a result, previous approaches do not estimate parameters by maximizing the exact GMCM likelihood but use a \textit{pseudo-likelihood} that has no guarantees of convergence or convergence to the correct parameters.
Further, this approach yields an inefficient optimization routine and biased parameter estimates \citep{bilgrau2016gmcm}.
The best previous approaches use variants of a Pseudo Expectation Maximization (PEM) approach for GMCM parameter estimation.

In this paper, we propose an approach, called AD-GMCM, that uses gradient-based optimization to obtain parameter estimates by maximizing the exact GMCM likelihood.
Our approach leverages 
Automatic Differentiation (AD) tools, 
that obviate the need to derive closed-form expressions of gradients and thereby facilitate Gradient Descent (GD) based inference\footnote{Most AD solvers use minimization as the canonical problem and ML estimation is done by Gradient Descent on the negative loglikelihood. Throughout the paper, we use the term {\it Gradient Descent} for maximization problems as well, instead of Gradient Ascent, with the assumption that the sign of the objective function is changed during optimization.}
 in complex models such as GMCM.
The computational efficiency and numerical accuracy offered by AD has been successfully leveraged in training deep neural networks.
They have also been used for optimization in some statistical models, e.g., %
non-linear random effects models \citep{fournier2012ad,skaug2006automatic,skaug2002automatic}, Bayesian models \citep{kucukelbir2017automatic} and mixture models \citep{maclaurin2015autograd}. 

Our extensive experiments on synthetic and real data show that AD-GMCM results in better GMCM likelihood and more accurate parameter estimates than PEM.
As a result, AD-GMCM also yields better performance in clustering of non-Gaussian data and in reproducibility analysis.
We theoretically analyze the conditions for monotonic increase in GMCM likelihood in the iterative procedures of PEM and AD-GMCM.
The advantages of an AD-based approach, to address problems related to monotonic increase of likelihood and parameter identifiability are discussed.
We also analyze two well-known cases of degeneracy of maximum likelihood in GMM that can lead to spurious clustering solutions.
Our analysis shows that surprisingly GMCM is not affected in one of the cases.

The rest of the paper is organized as follows.
We begin with an overview of GMM, Copulas, GMCM and its application in Clustering and Reproducibility Analysis (\S \ref{sec:background}).
We also provide a brief description of Automatic Differentiation and its advantages. We then describe previous approaches for GMCM parameter inference, followed by our proposed method, AD-GMCM (\S \ref{sec:paraminf}).
Theoretical considerations of monotonicity,  identifiability and degeneracy are discussed next (\S \ref{sec:theory}).
This is followed by
simulation studies (\S \ref{sec:sim}), experiments on real datasets (\S \ref{sec:real}) and our concluding discussion (\S \ref{sec:conclusion}).

\section{Background}
\label{sec:background}

Consider $n$ i.i.d. instances of $p$-dimensional data, $ [x_{i}^j]_{n \times p} = (\mathbf{X}^1,\ldots,\mathbf{X}^p) = (\mathbf{X}_1,\ldots,\mathbf{X}_n) $, where 
$i$ denotes the observation, and $j$ denotes the dimension and
$x_i^j$ denotes the $j$-th dimension value of the $i$-th observation.
Single superscript denotes the index along the dimensions and single subscript denotes the index along the observations, unless specified otherwise.
Appendix A contains the list of all symbols and notations used in the paper for reference.

\subsection{Gaussian Mixture Model (GMM)}
Let $\phi$ denote the multivariate normal distribution.
The Probability Density Function (PDF) of a $K$-component GMM is given by $f_{\text{GMM}}(\bx;\btheta) = \sum_{k=1}^K \pi_k \phi(\bx;\bmu_k,\bSigma_k)$,
with mixing proportions $\pi_k>0$ such that $\sum_{k=1}^K\pi_k=1$, component-specific mean vectors, $\vecmu_k$, and covariance matrices, $\bSigma_k$.
We use $\btheta= (\pi_1,...\pi_K, \bmu_1,...,\bmu_K, \bSigma_1,...,\bSigma_K)$ to denote all the parameters.

\subsection{Copulas} 
Let $f$ be the PDF of the distribution from which $\bX_i$'s are sampled. Let $F_j$ denote the marginal Cumulative Distribution Function (CDF) of $\mathbf{X}^j$ along the $j^{th}$ dimension.
A CDF transformation, $\mathbf{U}^j = F_j(\mathbf{X}^j)$, 
maps a random variable to a scalar that is uniformly distributed in $[0,1]$.
However, the joint distribution of all $p$ marginal CDFs is not uniform and is modeled by a copula, which  
is a multivariate distribution function $C: [0,1]^p  \rightarrow [0,1]$, 
defined on random variables $\mathbf{U}^j$.
Copulas uniquely characterize continuous joint distributions \citep{Sklar}:
for every joint distribution with continuous marginals, $F(\bX^1, \ldots, \bX^p)$, there exists a 
unique copula function such that $F(\bX^1, \ldots, \bX^p) = C(F_1(\bX^1), \ldots , F_p(\bX^p) )$; and the converse is also true.
It can be shown that 
the corresponding joint density is given by the product of the individual marginal densities $f_j$ and the \emph{copula density} $c$:
\begin{equation}\label{joint}
 f(\bx) = c(F_1(\bx^1), \ldots, F_p(\bx^p)) \,\, \Pi^p_{j=1}f_j(\bx^j).
\end{equation}

\begin{figure*}
    \centering
    \includegraphics[width=\textwidth]{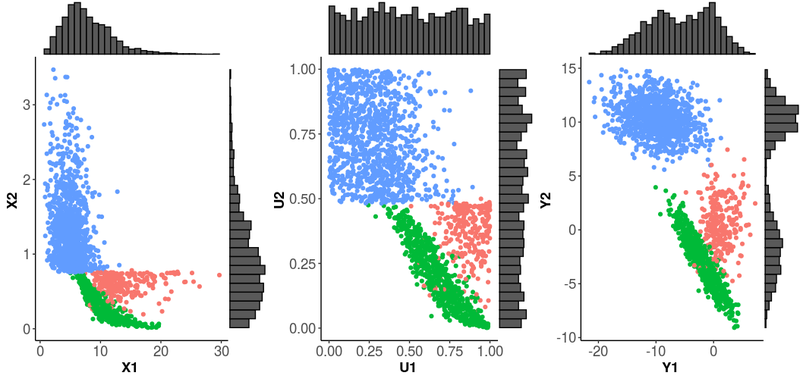}
    \caption{ Left: Observed data with the marginals $X1$ and $X2$ following Gamma and Weibull distributions respectively, 3 clusters are indicated by their respective colors(visually each of them appear to be non-Gaussian); 
    Center: scaled ranks with uniform marginals;
    Right: Latent observations where the individual marginals follow a Gaussian mixture model.
    }
    \label{fig:gmcm_illustration}
\end{figure*}

Eq. \ref{joint} shows how copulas enable flexible constructions of multivariate densities by decoupling the specification of marginals ($f_j$) and dependence structure ($c$), thus allowing us to choose each parametric family independently from each other (as shown in figure \ref{fig:gmcm_illustration}).

Eq. \ref{copula} (derived from (\ref{joint})), illustrates
how  copula families can be defined by the choice of the joint density $f$ that determines the dependence structure.
\begin{equation}\label{copula}
c(\bU^1,\ldots,\bU^p) = \frac{ f(\bx) } {\Pi^p_{j=1}f_j(\bx^j)}.
\end{equation}

For instance, the Gaussian copula density, $c_{\phi}$, is defined
by choosing the multivariate normal density $\phi$ with normal marginal densities $\phi_j$ in Eq. \ref{copula}: $c_{\phi} = \frac{ \phi(\vecx) } {\Pi^p_{j=1}\phi_j(\vecx^j)}$.
Copula-based models can also be viewed as generative models defined on the CDF-transformed data, $\mathbf{U}^j = F_j(\mathbf{X}^j)$ (also called pseudo or latent observations).
The generative model for the Gaussian copula is defined through a Gaussian distribution on the latent CDF transformations \citep{hoff2007extending}:
$
\mathbf{X}^{j} = F^{-1}_j (\mathbf{U}^{j});\,\,\,
\mathbf{U}^{j} = \Phi_j(\mathbf{Y}^{j});\,\,\,
\mathbf{Y} \sim \Phi(\vecmu,\bSigma)
$
where 
$\Phi_j$ and $\phi_j$ denote the $j^{th}$ marginal CDF and PDF respectively of the multivariate normal distribution with PDF $\phi$, CDF $\Phi$, mean $\vecmu$ and covariance matrix $\bSigma$.
We refer the reader to \cite{joe2014dependence} for a detailed account of copulas.

\subsection{Gaussian Mixture Copula Model (GMCM)}
In GMCM, the dependence is obtained from a GMM.
Let $\Psi_{j\theta}$,
and $\psi_{j\theta}$ denote the $j^{th}$ marginal CDF and PDF respectively, of the GMM $f_{\text{GMM}}(. ;\btheta)$. 
Using Eq. \ref{copula}, we obtain the 
GMCM copula density:
\begin{equation}\label{GMC}
c_{f}(\mathbf{U};\btheta) = 
\frac{ f_{\text{GMM}}(\Psi_{j\theta}^{-1} (\mathbf{U}))}{\prod_{j=1}^p \psi_j( \Psi_{j\theta}^{-1} (\mathbf{U}^j))} 
\end{equation}

The generative model of GMCM
is specified by:
\begin{equation}\label{GMCM}
\begin{split}
\mathbf{X}^j = F_j^{-1}(\mathbf{U}^j);\,\, \mathbf{U}^j = \Psi_j(\mathbf{Y}^j);\,\,  \mathbf{Y^j} \sim \psi_{j\theta} %
.
\end{split}
\end{equation}
Note that $\mathbf{U}^j = F_j(\mathbf{X}^j)= \Psi_j(\mathbf{Y}^j)$ and so, $\mathbf{Y}^j = \Psi_{j\theta}^{-1}(\mathbf{U}^j)$, 
where $\Psi_{j\theta}^{-1}$ denotes the inverse CDF of the latent GMM.
Figure \ref{fig:gmcm_illustration}  illustrates the GMCM generative process. There are 3 clusters which are indicated by their respective colors. The marginals of the actual observations (left) are Weibull and Gamma univariate distributions. Visual inspection suggests that none of these clusters are component-wise Gaussian. Scaling the observations by their ranks give uniform marginals (center). The pseudo (latent) observations (computed using the GMM inverse CDF) follow a GMM (right).   

The likelihood of $n$ i.i.d. samples from GMCM can be expressed in terms of the latent variable $\mathbf{Y}$:
\begin{equation}\label{eq:exactGMCM}
\mathcal{L} = \prod_{i=1}^n \frac{\sum_{k=1}^K \pi_k\phi(\vecy_i\mid\bmu_k,\bSigma_k)}{\prod_{j=1}^p \sum_{k} \psi_j(y_{i}^j \mid\mu_k^j,\Sigma_{kjj} ))}
\end{equation}

Note the difference between GMCM and the mixture of copulas model of \cite{kosmidis2016model}. The latter is a finite mixture model where each component density is defined using a copula (Eq. \ref{joint}). This  allows construction of new mixture models through the choice of arbitrary copula density and marginals in each component.
In contrast, GMCM, is a specific copula model, wherein the distribution of the CDF-transformed data ($\mathbf{U}_j$) is modeled with a GMM. Like any copula model, each marginal density can be independently specified. 

\subsection{Clustering and Reproducibility Analysis with GMCM}
The GMCM model has been found to be effective for clustering and reproducibility analysis.
Cluster labels $l \in \{1,\ldots,K\}$
can be obtained
by fitting GMCM to the data and obtaining
the maximum apriori estimate (MAP) $\argmax_k P(l = k| \btheta, \mathbf{X})$. 

In reproducibility analysis, the input $n \times p$ data matrix consists of test statistics or p-values from $p$ experiments investigating the same null hypothesis for $n$ subjects.
The aim is to identify a group of subjects that are commonly significant across experiments.
\cite{li2011reproducibility} proposed a special case of GMCM to solve this problem, which used a GMM ($f_{\text{GMM}}$) with 2 components corresponding to subjects satisfying the null and alternate hypothesis respectively.  Further, they impose constraints on the parameters such that in one component
the latent variable follows a $p$-dimensional standard multivariate normal distribution (component 1) and in the other component the latent variable follows a multivariate normal distribution with equal means and compound symmetry covariance structure (component 2). 
Component 1 is used to identify subjects for which the null hypothesis is accepted while the latter component is used to identify subjects for which the null hypothesis can be rejected.

With these constraints, the mean vectors are:
\begin{align}
\label{eq:repro1}
    \boldsymbol{\mu}_{1} & =\mathbf{0}_{p \times 1}=(0,0, \ldots, 0)^{\top} \\
\boldsymbol{\mu}_{2} & =\mathbf{1}_{p \times 1} \mu=(\mu, \mu, \ldots, \mu)^{\top}, \quad \mu>0 \nonumber
\end{align}
and the covariance matrices, 
$\bSigma_1$ is the identity matrix and $\bSigma_2$ has an equi-covariance structure:
\begin{equation}
\label{eq:repro2}
\mathbf{\Sigma}_{1}=\boldsymbol{I}_{p \times p}=\left[\begin{array}{ccc}
1 & 0 & \cdots \\
0 & 1 & \cdots \\
\vdots & \vdots & \ddots
\end{array}\right], 
\bSigma_{2}=\left[\begin{array}{ccc}
\sigma^{2} & \rho \sigma^{2} & \cdots \\
\rho \sigma^{2} & \sigma^{2} & \cdots \\
\vdots & \vdots & \ddots
\end{array}\right]\end{equation}
where, $\rho \in\left[-(p-1)^{-1}, 1\right] $ to ensure the positive definiteness of the matrix. 
Let $\alpha_1$ and $\alpha_2 = 1 - \alpha_1$ be the mixing proportions of the two components.
This special case of GMCM is parameterized by $\btheta = (\alpha_1, \mu, \sigma, \rho)$.

\cite{li2011reproducibility} define the \textit{local irreproducibility rate (idr)} of the $i$-th subject as $P(l \in \{1,2\}| \bu_i, \btheta)$ which coincides with the MAP for the general GMCM given above. 
This is similar to the local false discovery rate of  \cite{efron2004large,efron2007size}. Additionally, \cite{li2011reproducibility} also define an \textit{adjusted IDR} as $\operatorname{IDR}(\alpha)=P\left(l =1 \mid \boldsymbol{u}_i \in I_{\alpha}, \boldsymbol{\theta}\right)$, where $I_{\alpha}=\{\boldsymbol{u}_i \mid$ idr $(\boldsymbol{u}_i)<\alpha\},$ i.e., the probability of a gene being non-reproducible while in the rejection region. The adjusted IDR $(\alpha)$ relates to idr in the same manner as marginal false discovery rate (mFDR) relates to the local false discovery rate (lFDR) \citep{bilgrau2016gmcm}.
The signals belonging to component 1 are considered spurious and those corresponding to component 2 are considered genuine.  

\subsection{Automatic Differentiation (AD)}

In traditional gradient-based inference, gradients are required in closed form which becomes laborious or intractable to derive as the model complexity increases, such as in GMCM. 
To automate the computation of derivatives three classes of techniques have been developed:
(a) Finite Differentiation (FD) (b) Symbolic Differentiation (SD) and (c) Automatic or Algorithmic Differentiation (AD).
FD is easy to implement but is slow at high dimensions and often results in floating point errors.
SD gives exact symbolic expressions of derivatives but the run time, computational complexity and memory requirements do not scale with increasing number of parameters. %
AD overcomes these limitations of FD and SD
and provides efficient and accurate numerical evaluation of derivatives without requiring closed form expressions. 

The numerical computation of a function can be decomposed into a finite set of elementary operations. 
These operations most commonly include arithmetic operations and transcendental function evaluations.
The key idea of AD is to numerically compute the derivative of a function by combining the derivatives of the elementary operations through the systematic application of the chain rule of differential calculus.
The efficiency of the computation is improved by storing evaluated values of intermediate sub-expressions that may be re-used.
Backpropagation, used to train neural networks, is a specific form of AD which is more widely applicable.
We refer the reader to recent surveys \citep{baydin2018automatic,margossian2019review} for more details. 
In \ref{app:AD} we provide an illustrative example of AD and a comparison of runtime between AD and SD.

Efficient implementations of AD are available in several programming languages and frameworks, e.g. Python \citep{maclaurin2015autograd}, R \citep{pav2016madness}, PyTorch \citep{paszke2017automatic} and Stan \citep{carpenter2015stan}.
Many first and second order gradient-based optimization algorithms are implemented in these libraries.
In this paper, we use Adam, a first order method that computes individual adaptive learning rates for different parameters from estimates of first and second moments of the gradients \citep{kingma2014adam}.

\section {GMCM Parameter Inference}
\label{sec:paraminf}

Standard Maximum Likelihood (ML) inference requires specifying a parametric family for each marginal density and inferring the parameters simultaneously with the copula parameters which is computationally expensive for even moderate dimensions.
The two--step IFM procedure \citep{joe2014dependence} requires fitting a marginal along each dimension and then using the CDF transformation to obtain the input data (in the range $[0,1]$) for an ML estimation of copula parameters.
In reality, marginals are usually not known and may be difficult to estimate correctly. Also, we may be interested solely in the dependence structure and may want to circumvent marginal parameter estimation.
In such cases, a \textit{semiparametric approach} is to use rank--transformed scaled empirical marginals to 
estimate the copula parameters.

In the rank transformation, each data element $x_{i}^j$, 
of the column vector $\bX^j$ is transformed to $U_i^j = \frac{1}{n} \sum_{i=1}^n \mathbf{1}(X_{i}^j\leq x_{i}^j)$, the scaled rank of $x_{i}^j$ in all the observations of the feature, $X_j$. 
The scaled rank transformed features lie in the range $[0,1]$  and can be used directly to estimate the copula parameters.
Such a semi-parametric approach was found to be effective for clustering using GMCM \citep{compstat,rajan2016dependency,kasa2020gaussian} as well as for reproducibility analysis \citep{li2011reproducibility}.
Semi-parametric inference of GMCM 
still requires maximizing the likelihood in Eq. \ref{eq:exactGMCM} which is non-trivial due to the non-convex nature of the likelihood and the lack of closed form expressions for computing the $\by^j = \bPsi^{-1}_j (\bu^j)$ as described below. 

\subsection{Previous Approaches}
\label{sec:prev}

Maximizing the exact copula likelihood $\mathcal{C}$ (Eq. \ref{eq:exactGMCM}) is difficult and most previous approaches have instead obtained GMCM parameter estimates using the pseudo-likelihood:
\begin{equation}\label{pseudo}
\mathcal{L}_p = \prod_{i=1}^n \sum_{k=1}^K \pi_k\phi(\vecy_i\mid\bmu_k,\bSigma_k)
\end{equation}
Such pseudo-likelihood based estimates have also been used for the Gaussian copula \citep{hoff2007extending}.

The similarity of the pseudo-likelihood to GMM likelihood naturally suggests an EM-based estimation approach.
However, unlike the case of GMM estimation (directly from data), the pseudo-likelihood is in terms the inverse CDF values (also called pseudo-observations) $\by^j = \bPsi^{-1}_j (\bu^j)$ which are not constant across the EM iterations.
All the previous works on GMCM \citep{tewari2011parametric,li2011reproducibility,compstat,bilgrau2016gmcm,kasa2020gaussian} propose variants of a Pseudo EM (PEM) approach
wherein the algorithm iteratively alternates between estimating the latent observations $\by^j$ and then updating 
$\btheta$ using E and M steps.  

The inverse CDF $\bPsi^{-1}_j (\bu^j)$ does not have a known closed-form and so, estimation procedures have been proposed, broadly based on two strategies: (i) grid search and interpolation and (ii) approximation of the Gaussian CDF.

The grid search procedure involves computation of the the mixture CDF $\bPsi_{j}$ values with the current parameter estimates $\btheta$.
The monotonicity of $\bPsi_{j}$ is used to compute the inverse function by reflection around the identity line.
For instance, \cite{bilgrau2016gmcm} use 
1000 function evaluations of each component along each dimension $j$, which are spread equidistantly in the interval $\mu_k^j \pm 5 \sqrt{\Sigma_{kjj}}$. %
The mixture CDF $\bPsi_{j}$ at these 1000 points is evaluated by using the approximation of the error function erf $(x) \approx 1-\left(a_{1} t+a_{2} t^{2}+a_{3} t^{3}\right) \exp \left(-x^{2}\right)$ where $t=1 /(1+b x)$ and $a_{1}, a_{2}, a_{3},$ and $b$ are constants (given in  \cite{abramowitz1972handbook}). 
reproducibility analysis by \cite{li2011reproducibility} where  $\left[\min \left(-3, \mu -3\right), \max \left(3, \mu +3\right)\right]$ was used as the grid interval, with 1000 evaluations along each dimension $j$. Increasing the number of function evaluations and interval widths can lead to more accurate computation of the latent observations, but with significant increase in computational cost with increasing dimensionality incurred in each PEM iteration. 

The second, more scalable, approach used by \cite{compstat,kasa2020gaussian} is based on the first-order Taylor expansion of the Gaussian CDF, from which an approximate expression for the inverse CDF, in terms of the current parameter estimates $\btheta$ is derived. 
Additional conditions are imposed during the inverse CDF computation to ensure that the pseudo-likelihood does not decrease in each iteration.
While this approximation is easy to compute, its accuracy quickly deteriorates for those datapoints which fall outside the 1 standard deviation interval from the component means. To our knowledge, efficient computation of the CDF and the inverse CDF of univariate mixtures through function approximators remains an open problem.

In all the PEM based algorithms proposed previously, the wrong likelihood is optimized and only a psuedo maximum likelihood estimate is obtained. 
There is no guarantee of convergence or the convergence to the correct parameters.
Further, this approach yields an inefficient optimization routine and biased parameter estimates. This problem is discussed in detail by \cite{bilgrau2016gmcm}.

A few approaches to maximize the exact likelihood (Eq. \ref{eq:exactGMCM}) instead of the pseudo-likelihood have also been proposed.
\cite{tewari2011parametric} discuss an alternative Active-Set trust region algorithm based on numerical gradients.
\cite{bilgrau2016gmcm} implement and evaluate a suite of alternative numerical gradient-based optimization procedures, such as 
Nelder-Mead \citep{nelder1965simplex}, LBFGS \citep{byrd1995limited}, simulated annealing, and quasi-Newton methods.
Appropriate transformations are done to transform the problem into an unconstrained optimization problem to use these methods such as Cholesky decomposition combined with a log-transformation to ensure positive-definiteness of covariance matrices, Lagrange multipliers for mixture components for the general model; and log transform to ensure positive variance, affine and logit function composition to restrict the correlation values \citep{bilgrau2016gmcm}.
However, evaluating gradients numerically is prone to rounding off errors \citep{nocedal2006numerical}. Further,  \cite{bilgrau2016gmcm} report that PEM approach gives better clustering accuracy and is more robust compared to numerical gradient-based algorithms, such as BFGS, which often fail to run %
as the estimates of the covariance matrix become singular. 
In our experiments as well, we found that among the previous inference methods, PEM has more accurate and robust performance across wide range of datasets compared to previous methods that maximize the exact likelihood.

\subsection{Our Approach: AD-GMCM}

We propose the use of Automatic Differentition (AD) to obtain GMCM parameter estimates by maximizing the exact likelihood (Eq. \ref{eq:exactGMCM}).
The key advantage that AD offers in this case is that it can provide efficient and accurate numerical evaluation of the gradients without requiring them in closed form.
Further, the second-order Hessian matrix may also be evaluated using AD, 
enabling us to use methods with faster convergence.

To reformulate the problem into an unconstrained optimization problem, we apply the following transformations:
\begin{itemize}
\item 
To ensure Positive Definiteness (PD) of the estimated covariance matrices, 
we compute the gradients with respect to $\bV_k$, where $\bSigma_k = \bV_k \bV^{T}_k$. We first initialize the values of $\bV_k$ as identity matrices. Thereafter, updated values of $\bV_k$ can be computed using gradient descent, i.e., $\bV_k := \bV_k + \epsilon \times \frac{\partial \mathcal{L}}{\partial \bV_k}$. Here, $\epsilon$ is the learning rate. %
If the gradients are evaluated with respect to $\bSigma_k$ directly, there is no guarantee that updated $\bSigma_k = \bSigma_k + \epsilon \times \frac{\partial \mathcal{L}}{\partial \bSigma_k}$ will still remain PD. However, if gradients are evaluated with respect to $\bV_k$, no matter what the updated matrix $\bV_k$ is, by construction $\bSigma_k$ always remains PD. Alternatively, one could use a Cholesky decomposition for $\bSigma_k$ \citep{salakhutdinov2003optimization}.

\item
 To ensure that the mixture proportions of components add up to one, we use the logsumexp trick \cite{robert2014machine}. We start with unbounded $\alpha_k$'s as the log-proportions, i.e.,  $\log \pi_k = \alpha_k - {\log(\sum_{k^'} e^{\alpha_{k^'}})}$.  Note that, we need not impose any constraints on $\alpha_k$ as final computation of $\pi_k$ automatically leads to normalization, because $\pi_k = \frac{\alpha_k}{\sum_{k^'} e^{\alpha_{k^'} }}$. Therefore, we can update $\alpha_k := \alpha_k + \epsilon \times \frac{\partial L}{\partial \alpha_k}$ without any further need for Lagrange multipliers. 

\end{itemize}

Thus, our algorithm, called AD-GMCM, proceeds iteratively with two steps in each iteration:
\begin{enumerate}
\item 
Maximization: of the exact GMCM likelihood through gradient-based AD
\item
Reset $\by_j$: obtain the current estimates of inverse CDF values $\by_j$. 
\end{enumerate}

The complete algorithm is shown in Algorithm \ref{AutoGMCMalgo}.
In our experiments, we use  the grid search and interpolation method of \cite{bilgrau2016gmcm} to obtain the updated $\by_j$.

\begin{algorithm}[t]
\caption{AD-GMCM}
\label{AutoGMCMalgo}
\footnotesize
\begin{algorithmic}
\STATE \textbf{Input: }
Observed $n$ datapoints (as $n \times p$ dimensional matrix $X=(\vecx_1, \vecx_2,...,\vecx_n)$), number of clusters $K$ and the convergence threshold $\gamma$
 \STATE \textbf{Initialize:} 
Set $\btheta^{(0)}$ using random start or \textit{K}-means clustering under the constraints that $\hat \pi_g^{(0)} >0$, $\sum_{k=1}^K \hat \pi^{(0)}_k =1$ and $\hat \bSigma^{(0)}_k$ is positive definite.
Set $u_{i}^j=\mathbf{\tilde{F}_j}(x_{i}^j)$ (percentile ranks).

\REPEAT
\STATE \textbf{Reset $\vecy$}:\\ 

$$y_{i}^{j^{(t+1)}} = \Psi_{j\theta^{(t)}}^{-1}(\mathbf{U}^j_i)$$
\STATE \textbf{Maximization:}
\resizebox{\linewidth}{!}{
  \begin{minipage}{\linewidth}
  \begin{align*}
\hat{\alpha}^{(t+1)}_k := \hat{\alpha}^{(t)}_k + \epsilon \frac{\partial \mathcal{L} }{\partial \alpha_k} ; \;\;
\hat{\bmu}^{(t+1)}_k & := \hat{\bmu}^{(t+1)}_k + \epsilon  \frac{\partial \mathcal{L}}{\partial \bmu_k} ;\;\;
\hat{\bV}^{(t+1)}_k  := \hat{\bV}^{(t)}_k + \epsilon \frac{\partial  \mathcal{L}}{\partial \bV_k} ; 
\;\\
\hat{\pi}^{(t+1)}_k  := \frac{ e^{\hat{\alpha}^{(t+1)}_k} }{\sum_{k^'} e^{\hat{\alpha}^{(t+1)}_{k^'}}} ;
& \;\;
\hat{\bSigma}^{(t+1)}_k := \hat{\bV}^{(t+1)}_k \hat{\bV}^{{(t+1)}^T}_k 
\end{align*}
  \end{minipage}
}

\UNTIL convergence criterion $|\mathcal{L}^{(t+1)} - \mathcal{L}^{(t)} | < \gamma $ is met.
\end{algorithmic}
\end{algorithm}

For Reproducibility Analysis, number of clusters ($K$) is set to 2 and $\mu_1, \Sigma_1$ are fixed as given in Eq. \ref{eq:repro1} and \ref{eq:repro2}.
Parameters $\alpha_1,\mu, \sigma$ are
obtained by maximizing the exact GMCM likelihood using Gradient Descent.
An additional constraint that $\rho \in\left[-(p-1)^{-1}, 1\right]$ is needed ensure that the covariance matrix is positive definite. This can be easily ensured by adding a penalty to the likelihood of the form ${\frac{(2\rho - (1+a))}{1-a}}^{2b}$ where $a = -(p-1)^{-1} $  and $b$ is large positive number. This penalty is close to zero when $\rho \in\left[-(p-1)^{-1}, 1\right]$ and blows up when $\rho$ is outside this range. %

\section{Theoretical considerations}
\label{sec:theory}

\subsection{Monotonicity}

It is well-known that the PEM algorithm (and its variants) does not ensure a monotonic increase in copula likelihood value over the iterations because the pseudo likelihood is  maximized \citep{kasa2020gaussian,bilgrau2016gmcm, compstat}. 
Despite optimizing the exact likelihood, AD-GMCM also does not ensure monotonicity. This is because there is no closed form expression for inverse CDF computation. Had there been a closed form expression, one could directly substitute $\Psi^{-1}(u)$ (in place of $\by$) into Eq. \ref{eq:exactGMCM} and proceed with evaluating the gradients. In order to tackle the lack of closed form for inverse CDF computaiton, both PEM and AD-GMCM have an additional `Reset y' step in addition to the maximization step. While the maximization step preserves the monotonicity, the `Reset y' step provides no such guarantees. 

For PEM, several modifications have been proposed in the `Reset y' step that ensure the monotonicity at the cost of deviating slightly from the actual pseudo likelihood \citep{compstat,kasa2020gaussian}. The key idea behind these modifications is to ensure there is no abnormal drop in the likelihood during the `Reset y' step. We derive similar conditions to ensure monotonicity for the exact likelihood in Lemma \ref{lemma:montonicity}. However, in practice, a simpler alternative in gradient-based solutions is to choose a small learning rate.
Empirically we find that it ensures  monotonic increase of copula likelihood during the iterations.

\begin{lemma}
\label{lemma:montonicity}
The copula likelihood monotonically increases with every iteration if the following conditions are ensured during the `Reset y' step in Algorithm \ref{AutoGMCMalgo}. 
\begin{enumerate}
    \item $|\by_i^{(t+1)} - \bmu_k^{(t+1)}| \succeq |\by_i^{(t)} - \bmu_k^{(t+1)}|$
    \item If ${y_i^j}^{(t+1)} - {y_i^j}^{(t)} \geq 0$ then ${y_i^j}^{(t+1)} \leq \frac{2\sum_k \mu_k^j}{k} - {y_i^j}^{(t)} $
\item If ${y_i^j}^{(t+1)} - {y_i^j}^{(t)} < 0$ then ${y_i^j}^{(t+1)} > \frac{2\sum_k \mu_k^j}{k} - {y_i^j}^{(t)} $
\end{enumerate}

\end{lemma}

The proof is given in  \ref{sec:proofmono}. As can be seen from Eq. \ref{eq:exactGMCM}, the copula likelihood is composed of the pseudo likelihood and marginal likelihood. Condition 1 ensures that marginal likelihood terms in the denominator decrease while conditions 2 and 3 ensure that pseudo likelihood term in the numerator increases.  A feasible point satisfying these conditions can be obtained by solving a straightforward linear programming problem. The conditions given in this lemma are less restrictive compared to those in \cite{compstat,kasa2020gaussian}.
Note that less restrictive conditions are crucial to ensure that we are indeed maximizing the copula likelihood.
For instance, consider the extreme case where we set $\by^{(t)} = \by^{(0)}$ for all $t\geq 1$. In such a case, even though the \textit{incorrect} evaluation of likelihood increases with every iteration, we are no longer maximizing the copula likelihood as the condition $\by = \Psi^{-1}(\bu)$ does not hold. Obtaining the least restrictive conditions for monotonicity remains an open problem. 

\subsection{Parameter unidentifiability}
\label{sec:param_unident}
The GMCM model suffers from unidentifiability of parameters as discussed in \cite{bilgrau2016gmcm}. This is because copulas are invariant to translation and scaling, and so they preserve only relative distances. We illustrate the identifiability issue using the following simulations. Consider the following 2-dimensional 3-component datasets where the first component is centered at the origin with a unit covariance matrix. The second component is varied as shown in table \ref{tab:ident_sim} to simulate different cluster separations and dependency structures.

\begin{table}[h!]
\centering
\begin{tabular}{|c|c|c|}
\hline
Set 1 &
$f_1$ = MVN $\biggl( \mu = (0,0) , \bSigma = \begin{pmatrix}
1 & 0 \\
0 & 1 
\end{pmatrix}    \biggr)$  & $f_2$ = MVN $ \biggl( \mu = (5,5) , \bSigma = \begin{pmatrix}
1 & 0 \\
0 & 1 
\end{pmatrix} \biggr) $   \\ \hline
Set 2 & $f_1$ = MVN  $ \biggl( \mu = (0,0) , \bSigma = \begin{pmatrix}
1 & 0 \\
0 & 1 
\end{pmatrix}  \biggr)$ &  $f_2$ = MVN $\biggl( \mu = (10,10) , \bSigma = \begin{pmatrix}
1 & 0 \\
0 & 1 
\end{pmatrix} \biggr)$  \\ \hline
Set 3 & $f_1$ = MVN $\biggl( \mu = (0,0) , \bSigma = \begin{pmatrix}
1 & 0 \\
0 & 1 
\end{pmatrix}  \biggr)$ & $f_2$ = MVN $\biggl( \mu = (5,5) , \bSigma = \begin{pmatrix}
1 & 0.5 \\
0.5 & 1 
\end{pmatrix}  \biggr)$ \\ \hline
\end{tabular}
\caption{Parameters of mixture component distributions used for illustrating unidentifiability of GMCM parameters.}
\label{tab:ident_sim}
\end{table}

\begin{figure}[!h]
    \centering
    \begin{minipage}{.995\textwidth}
        \centering
    \begin{subfigure}[b]{0.4\textwidth}
\includegraphics[width= \textwidth]{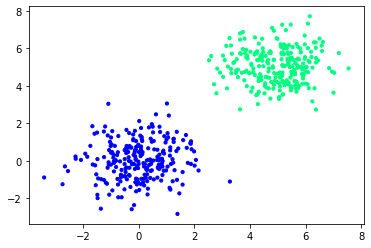}
        \caption{Set 1 - Observations}    
    \end{subfigure}%
    \begin{subfigure}[b]{0.4\textwidth}
\includegraphics[width=\textwidth]{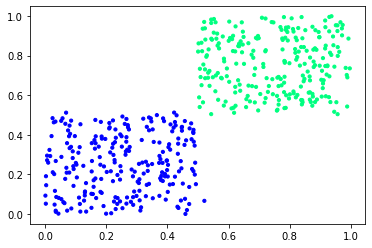}
        \caption{Set 1 - Scaled Rank Values}  
    \end{subfigure}

    \begin{subfigure}[b]{0.4\textwidth}
\includegraphics[width= \textwidth]{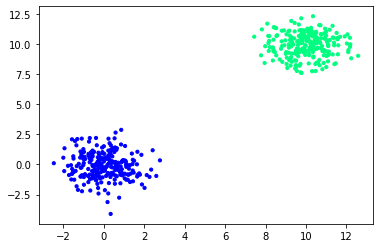}
        \caption{Set 2 - Observations}    
    \end{subfigure}%
    \begin{subfigure}[b]{0.4\textwidth}
\includegraphics[width=\textwidth]{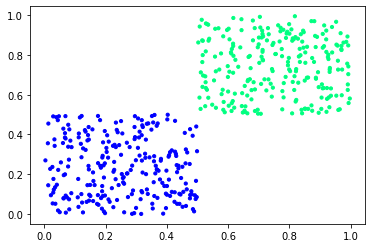}
        \caption{Set 2 - Scaled Rank Values}  
    \end{subfigure}

    \begin{subfigure}[b]{0.4\textwidth}
\includegraphics[width= \textwidth]{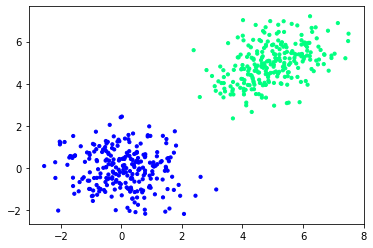}
        \caption{Set 3 - Observations}    
    \end{subfigure}%
    \begin{subfigure}[b]{0.4\textwidth}
\includegraphics[width=\textwidth]{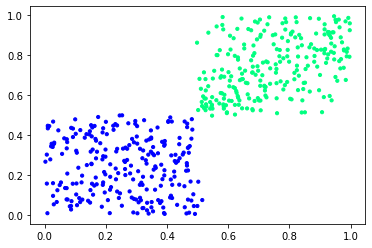}
        \caption{Set 3 - Scaled Rank Values}  
    \end{subfigure}

    \caption{Illustration of unidentifiability: Clusters in Set-1 and Set-2 have the same rank distribution despite having different distributions in the observation space. Only the correlation and mixture weight parameters of GMCM can be reliably estimated.}
    \label{fig:identifiability_illustration}
    \end{minipage}
    \end{figure}

As seen in figure \ref{fig:identifiability_illustration}, the scaled rank distributions for 
Set 1 and Set 2 are almost identical despite having completely different component means. However, when a correlation (0.5 in $f_2$) is introduced in Set 3, we can observe that the scaled rank distributions are no longer identical. This illustrates that while we cannot reliably estimate the component means and covariances, we can still estimate the component correlations and component weights. In order to mitigate the effect of unidentifiability of component means and weights, \cite{bilgrau2016gmcm} suggest anchoring one of the components to have a mean of $(0,0,\dots,0)$ and covariance matrix $\mathbb{I}$. Implementing this anchoring solution in PEM methodology requires deriving new update expressions; however, it can be done without additional effort in AD-GMCM by simply changing the differentiable parameters of the model.

\subsection{Degeneracy in GMCM}
Degeneracy of maximum likelihood in Gaussian mixture models have been well studied. The likelihood of GMM with unrestricted covariance matrices is unbounded \citep{day1969estimating,ingrassia2004likelihood,Chen2009,ingrassia2011degeneracy}. This leads to degeneracy when the component covariance matrices become singular. 
As a result, the corresponding clustering solution has a cluster with very few closely located data points.
In the following, we illustrate (a) a well-known cause of degeneracy which is present is GMM but \textit{absent} in GMCM and (b) an atypical cause of degeneracy present in both GMCM and GMM.

\subsubsection{Lack of Degeneracy when variance tends to zero}
Degeneracy occurs in GMM during ML estimation primarily due to the fact that one can always choose to have the mean of a component match with a datapoint and make the covariance of that component arbitrarily small \citep{McLa:Peel:fini:2000}. 
Interestingly, we find that such unboundedness in likelihood disappears for GMCM. To illustrate this, consider a 2-component 2-dimensional Gaussian mixture in the latent space with means $\bmu_1 = (\mu_1^{1},\mu_1^{2}),\bmu_2 = (\mu_2^{1},\mu_2^{2}) $, and with precision matrices $\bSigma_1^{-1} = \begin{pmatrix} \frac{1}{\sigma_{11}^2 } & 0 \\ 0 & \frac{1}{\sigma_{12}^2} \end{pmatrix}, \bSigma_2^{-1} = \begin{pmatrix} \frac{1}{\sigma_{21}^2 } & 0 \\ 0 & \frac{1}{\sigma_{22}^2} \end{pmatrix}$. First, we consider the GMM case where there are only two observations and both of them coincide with the component means i.e. $(\by_1,\by_2) = (\bmu_1,\bmu_2)$. The log-likelihood is given by 
\begin{align}
   \underbrace{ \ln \left(  \frac{\pi_1}{\sigma_{11}\sigma_{12} } +  \frac{\pi_2}{\sigma_{21}\sigma_{22} } e^{-\left( \frac{(\mu_1^1 -\mu_2^1)^2}{2\sigma_{21}^2} + \frac{(\mu_1^2 -\mu_2^2)^2}{2\sigma_{22}^2} \right)} \right) }_{A} +  \ln \left(  \frac{\pi_2}{\sigma_{21}\sigma_{22} } +  \frac{\pi_1}{\sigma_{11}\sigma_{12} } e^{-\left( \frac{(\mu_1^1 -\mu_2^1)^2}{2\sigma_{11}^2} + \frac{(\mu_1^2 -\mu_2^2)^2}{2\sigma_{12}^2} \right)} \right) \nonumber
\end{align}

As $\sigma_{11} \to 0$, the first term $A$ in the above expression blows up, leading to unbounded likelihood. Note that the term $\frac{\pi_1}{\sigma_{11}}e^{-\frac{x^2}{\sigma_{11}^2}}$ goes to zero as $\sigma_{11} \to 0$ for all nonzero x.  On the other hand, consider the expression of GMCM log likelihood for the same datapoints. 

\begin{align*}
   \underbrace{ \ln \left(  \frac{\pi_1}{\sigma_{11}\sigma_{12} } +  \frac{\pi_2}{\sigma_{21}\sigma_{22} } e^{-\left( \frac{(\mu_1^1 -\mu_2^1)^2}{2\sigma_{21}^2} + \frac{(\mu_1^2 -\mu_2^2)^2}{2\sigma_{22}^2} \right)} \right)}_{B} 
     - \underbrace{\ln \left( \frac{\pi_1}{\sigma_{11}} + \frac{\pi_2}{\sigma_{21}} e^{-\frac{ (\mu_{1}^1 - \mu_2^1)^2  }{2\sigma_{21}^2 }} \right)}_{C} - \ln \left( \frac{\pi_1}{\sigma_{12}} + \frac{\pi_2}{\sigma_{22}} e^{-\frac{ (\mu_{1}^2 - \mu_2^2)^2  }{2\sigma_{22}^2 }} \right) 
    \\    + \ln \left(  \frac{\pi_2}{\sigma_{21}\sigma_{22} } +  \frac{\pi_1}{\sigma_{11}\sigma_{12} } e^{-\left( \frac{(\mu_1^1 -\mu_2^1)^2}{2\sigma_{11}^2} + \frac{(\mu_1^2 -\mu_2^2)^2}{2\sigma_{12}^2} \right)} \right) 
    -\ln \left( \frac{\pi_2}{\sigma_{21}} + \frac{\pi_1}{\sigma_{11}} e^{-\frac{ (\mu_{1}^1 - \mu_2^1)^2  }{2\sigma_{11}^2 }} \right) - \ln \left( \frac{\pi_2}{\sigma_{22}} + \frac{\pi_1}{\sigma_{12}} e^{-\frac{ (\mu_{1}^2 - \mu_2^2)^2  }{2\sigma_{12}^2 }} \right) 
\end{align*}

As $\sigma_{11} \to 0$, both the first term $B$ and second term $C$ in the above expression blow up, leading to a finite limit (evaluated using L-Hospital's rule). Hence, the degeneracy that plagues ML estimation in GMM is absent in GMCM.

\subsubsection{Degeneracy when correlation tends to 1}
Similar to GMM, there does exist a degeneracy in the likelihood when individual correlations go to one. To illustrate this, consider the same setting as above but with precision matrices $\bSigma_1^{-1} = \frac{1}{(1-\rho_1^2)}\begin{pmatrix} \frac{1}{\sigma_{11}^2 } & -\frac{\rho_1}{\sigma_{11}\sigma_{12}} \\ -\frac{\rho_1}{\sigma_{11}\sigma_{12}} & \frac{1}{\sigma_{12}^2} \end{pmatrix}, \bSigma_2^{-1} = \frac{1}{(1-\rho_2^2)} \begin{pmatrix} \frac{1}{\sigma_{21}^2 } & -\frac{\rho_2}{\sigma_{21}\sigma_{22}} \\ -\frac{\rho_2}{\sigma_{21}\sigma_{22}} & \frac{1}{\sigma_{22}^2} \end{pmatrix}$

The log likelihood can be calculated as 

\begin{align*}
    \underbrace{\ln \left(  \frac{\pi_1}{ \sqrt{1-\rho_1^2}\sigma_{11}\sigma_{12} } +  \frac{\pi_2}{\sqrt{1-\rho_2^2}\sigma_{21}\sigma_{22} } e^{-\frac{1}{(1-\rho_2^2)}\left( \frac{(\mu_1^1 -\mu_2^1)^2}{2\sigma_{21}^2} + \frac{(\mu_1^2 -\mu_2^2)^2}{2\sigma_{22}^2} -\frac{\rho_2(\mu_1^1 -\mu_1^2)(\mu_1^2 -\mu_2^2)}{\sigma_{21} \sigma_{22} } \right)}    \right)}_{ D} \\
     -\ln \left( \frac{\pi_1}{\sigma_{11}} + \frac{\pi_2}{\sigma_{21}} e^{-\frac{ (\mu_{1}^1 - \mu_2^1)^2  }{2\sigma_{21}^2 }} \right) - \ln \left( \frac{\pi_1}{\sigma_{12}} + \frac{\pi_2}{\sigma_{22}} e^{-\frac{ (\mu_{1}^2 - \mu_2^2)^2  }{2\sigma_{22}^2 }} \right) 
    \\    + \ln \left(  \frac{\pi_2}{ \sqrt{1-\rho_2^2} \sigma_{21}\sigma_{22} } +  \frac{\pi_1}{ \sqrt{1-\rho_1^2} \sigma_{11}\sigma_{12} } e^{ - \frac{1}{(1-\rho_1^2)}  \left( \frac{(\mu_1^1 -\mu_2^1)^2}{2\sigma_{11}^2} + \frac{(\mu_1^2 -\mu_2^2)^2}{2\sigma_{12}^2} -\frac{\rho_1(\mu_1^1 -\mu_1^2)(\mu_1^2 -\mu_2^2)}{\sigma_{11} \sigma_{12} } \right)} \right) \\
    -\ln \left( \frac{\pi_2}{\sigma_{21}} + \frac{\pi_1}{\sigma_{11}} e^{-\frac{ (\mu_{1}^1 - \mu_2^1)^2  }{2\sigma_{11}^2 }} \right) - \ln \left( \frac{\pi_2}{\sigma_{22}} + \frac{\pi_1}{\sigma_{12}} e^{-\frac{ (\mu_{1}^2 - \mu_2^2)^2  }{2\sigma_{12}^2 }} \right) 
\end{align*}

As $\rho_1 \to 1$, only the first term D in the above expression goes to infinity. Note that the term $$\frac{\pi_1}{ \sqrt{1-\rho_1^2} \sigma_{11}\sigma_{12} } e^{-\frac{1}{(1-\rho_1^2)}\left( \frac{(\mu_1^1 -\mu_2^1)^2}{2\sigma_{11}^2} + \frac{(\mu_1^2 -\mu_2^2)^2}{2\sigma_{12}^2} -\frac{\rho_1(\mu_1^1 -\mu_1^2)(\mu_1^2 -\mu_2^2)}{\sigma_{11} \sigma_{12} } \right)}$$ goes to zero as $\rho_1 \to 1$. Hence, as the individual correlations approach one, and there is a degeneracy in likelihood. 

As seen from the above discussion, degenerate clustering solutions for GMCM usually occur at the boundary of parameter space (where individual correlations are one), similar to those found in GMM \citep{McLa:Peel:fini:2000}.

\section{Simulation Study}
\label{sec:sim}

We compare the performance of our proposed AD-GMCM with that of PEM using simulated datasets in three tasks: (1) GMCM Parameter Estimation, (2) Clustering, and (3)  
Reproducibility Analysis.

We use the implementation of PEM from \cite{bilgrau2016gmcm}, which has been compared with other methods of GMCM parameter estimation, as well as GMCM-based clustering and reproducibility analysis.
As reported by \cite{bilgrau2016gmcm}, we also find that other gradient-based optimization methods such as Nelder-Mead, LBFGS often fail due to singular covariance matrix estimates and are too slow to converge.
For a fair comparison, we use the same method of estimating the inverse CDF values in the `Reset y' step of Algorithm \ref{AutoGMCMalgo} as that used in PEM based on grid search and interpolation (described in section \ref{sec:prev}).

\subsection{Parameter Estimation}

We conduct two sets of simulations with 30 datasets each where we simulate 2-component 2-dimensional mixtures with varying cluster separation. For both sets of experiments, we set $\bmu_1 = (0,0)$,  $\bSigma_1 = \begin{pmatrix}
1 & -0.5 \\
-0.5 & 1 
\end{pmatrix}, \bSigma_2 = \begin{pmatrix}
1 & 0.5 \\
0.5 & 1 
\end{pmatrix}$. For the setting where comonents are not well-separated, we set $\bmu_2= (1,1)$ and for the well separated case, we set $\bmu_2 = (3,3)$.

Table \ref{table:param_ident_sim} shows the difference between true values and the estimates of mixture proportions and correlations obtained by PEM and AD-GMCM.
As discussed in section \ref{sec:param_unident}, other parameters are unidentifiable.
We observe that the accuracy of parameter estimation of both PEM and AD-GMCM improves with increase in cluster separation.  AD-GMCM, by virtue of maximizing the exact likelihood, leads to a better estimation of parameters in both the cases.  

\begin{table*}[!htb]
\centering
\begin{tabular}{|c|p{1cm} p{1cm}p{1cm}p{1cm}|p{1cm}p{1cm}p{1cm}p{1cm}| } 
 \hline
  & \multicolumn{4}{c |}{Not well-separated}  & \multicolumn{4}{c|}{Well-separated}   \\ \hline
 Method & $\pi_1$ & $\pi_2$ & $\rho_1$ & $\rho_2$   & $\pi_1$ & $\pi_2$ & $\rho_1$ & $\rho_2$  \\ \hline
 
 PEM & 0.031 \newline (0.122) & -0.032 \newline (0.122) & -0.163 \newline (0.240) & -0.16 \newline (0.262) & -0.012 \newline (0.075) & 0.099 \newline (0.075) & -0.027 \newline (0.158) & 0.001 \newline (0.120)  \\ \hline 

AD-GMCM & 0.025 \newline (0.112) & -0.026 \newline (0.112) & -0.108 \newline (0.262) & -0.084 \newline (0.245) & -0.009 \newline (0.073) & 0.008 \newline (0.073) & -0.012 \newline (0.192) & 0.050 \newline (0.118) \\ \hline

\end{tabular}
\captionof{table}{Mean difference (with standard deviation) of AD-GMCM and PEM parameter estimates from true values, over 30 datasets.}
\label{table:param_ident_sim}
\end{table*}

\subsection{Clustering Accuracy}

We compare clustering performance, measured by the Adjusted Rand Index (ARI), of AD-GMCM and PEM on two sets of simulations.
In the first set, data is simulated from a GMCM.
To test the clustering performance of the methods, when there are deviations from the assumed model, we evaluate the performance on non-Gaussian mixtures that are not generated through GMCM.

\subsubsection{GMCM Data}
We use the GMCM package \citep{bilgrau2016gmcm} to simulate datasets from the GMCM model. The GMCM package provides a  function to simulate datasets for input  values of $K$ and $p$. The mean and covariance parameters are randomly chosen by the function.
This allows us to test for varying settings of covariances structures, cluster separation, and unbalanced components.   We vary $K,p$ such that $K,p \in \{2,3,4\}$, for a total of nine combinations of $(K,p)$. For each combination of $K$ and $p$ values, we simulate 30 different datasets for a total of 270 datasets.

\begin{table}[h!]
\centering
\begin{tabular}{cc c cc c cc c ccc c cc}
\toprule
 $K$ & $p$   && \multicolumn{2}{c}{Ind. LL} &&  \multicolumn{2}{c}{Avg. LL} && \multicolumn{3}{c}{Ind. ARI} && \multicolumn{2}{c}{Avg. ARI} \\
 \cline{4-5}\cline{7-8}\cline{10-12}\cline{14-15}
&  && $>$ & $\leq$ && AD-GMCM & PEM  && $>$ & $=$ & $<$  && AD-GMCM & PEM \\  
\midrule
4 & 2    && 30 & 0  && 54.292 & 51.658   && 12 & 13   & 5         && 0.470 & 0.459        \\ 
4 & 4    &&  30  & 0   &&  235.640  & 232.322   && 10   & 19 & 1      &&  0.728  & 0.721     \\ 

4 & 3   &&  30 &  0  &&  149.183 & 138.306  &&  4 & 19   & 7   &&  0.687 &    0.663  \\

3 & 4   &&  30 &  0 &&  218.693 &  215.914  &&  3 & 24   & 3   &&  0.767 &    0.761  \\

3 & 3   &&  30 &  0 &&  135.437 &  133.545  &&  3 & 24   & 3   &&  0.766 &    0.748  \\

3 & 2   &&  30 &  0  &&  71.068 &  66.790 &&  5 & 19   & 6   &&  0.634 &    0.630  \\ 

2 & 2   &&  30 &  0 &&  49.830 &  46.815  &&  4 & 21   & 5   &&  0.685 &    0.666  \\ 

2 & 4   &&  30 &  0 &&  134.067 &  131.081  &&  3 & 24   & 3   &&  0.793 &    0.762  \\ 

2 & 3   &&  30 &  0 &&  104.855 &  99.063  &&  5 & 22   & 3   &&  0.761 &    0.678  \\ \hline
Total & && 270 & 0 & & & & &  49 & 185 & 36 & & &\\ 
\bottomrule
\end{tabular}
\caption{Log-likelihood (LL) and ARI for AD-GMCM vs PEM over 10 different simulation settings. Indivdual (Ind.) results show the number of datasets in which AD-GMCM obtains higher, equal or lower values compared to PEM. Average (Avg.) results are over 30 datasets in each setting.}
\label{tab:clus_accu}
\end{table}

Table \ref{tab:clus_accu} shows the number of times AD-GMCM obtains higher, equal and lower log-likelihood and ARI, compared to PEM for each setting
\footnote{Please see our note in \ref{sec:highLL} that explains why the log-likelihood values for GMCM are high.}
.
The average likelihood and ARI values, in each setting, are also shown. 
In all the 270 datasets, we observe that  AD-GMCM leads to a solution with a higher copula likelihood compared to PEM. 
The average ARI obtained by AD-GMCM is higher than that of PEM in 8 out of 10 settings.

\subsubsection{Other Non-Gaussian Mixtures}

We simulate 4 different sets (a total of 240 datasets) of data from non-Gaussian mixtures, similar to the evaluation in \cite{compstat}. 
Within each set, we further simulate two kinds of data  -- a highly correlated set and weakly correlated set, of 30 datasets each. All simulations contain  2-dimensional datapoints present in 3 clusters. 
Each datapoint is the product of a sample from a multivariate normal distribution (MVN) and another distribution as outlined in Table \ref{tab:non_gmcm_sim_details_1}. 
We chose the same MVN parameters for clusters 1 and 3, in order to study whether the algorithms can distinguish the two clusters. We get 8 different sets of simulations by varying the shape, location, and correlation parameters as given in table \ref{tab:non_gmcm_sim_details_2}.
 The odd numbered sets in the table are weakly correlated and the even numbered sets are strongly correlated. 

\begin{table}[h!]
\centering
\begin{tabular}{|l|}
\hline
$f_1$
= MVN $\bigg( \mu = (-5,-5) ,
 \bSigma
= \begin{pmatrix}
0.5 & \rho_1 \\
\rho_1 & 0.5 
\end{pmatrix} \bigg) \times $ Unif(0,1)  \\  \hline
$f_2$ 
= MVN $\bigg( \mu = (2,2)$ , $\bSigma = \begin{pmatrix}
1 & 0 \\
0 & 5 
\end{pmatrix} \bigg) \times$ Weibull(scale = sc_1, shape = sh_1)  \\ \hline
$f_3$ 
= MVN $ \bigg( \mu = (-5,-5)$ , $\bSigma = \begin{pmatrix}
0.5 & \rho_3 \\
\rho_3 & 0.5 
\end{pmatrix} \bigg) \times$ Gamma(shape = sh_2, scale = sc_2)  \\ \hline
\end{tabular}
\caption{Distributions used in simulation of 2-dimensional, 3-component non-Gaussian mixtures }

\label{tab:non_gmcm_sim_details_1}
\end{table}

\begin{table}[!h]
    \centering
    \begin{tabular}{|c|c|c|c|c|c|c|} \hline
       Set No. &   $\rho_1$  & $\rho_3$ & $sc_1$ & $sh_1$ & $sc_2$ & $sh_2$  \\ \hline
        1 & 0  & 0 & 1 & 2 & 1 & 2 \\ \hline
        2 & 0.45  & 0.45 & 1 & 2 & 1 & 2 \\ \hline
        3 & 0  & 0 & 1 & 2 & 2 & 2 \\ \hline
        4 & 0.45  & 0.45 & 1 & 2 & 2 & 2 \\ \hline
        5 & 0  & 0 & 2 & 2 & 1 & 2 \\ \hline
        6 & 0.45  & 0.45 & 2 & 2 & 1 & 2 \\ \hline
        7 & 0  & 0 & 2 & 2 & 2 & 2 \\ \hline
        8 & 0.45  & 0.45 & 2 & 2 & 2 & 2 \\ \hline
    \end{tabular}
    \caption{Shape, location and correlation parameters used to simulate data from non-Gaussian mixtures.}
    \label{tab:non_gmcm_sim_details_2}
\end{table}
\begin{table}[h!]
\centering
\begin{tabular}{c c cc c cc c ccc c cc}
\toprule
 Set No. && \multicolumn{2}{c}{Ind. LL} && \multicolumn{2}{c}{Avg. LL} && \multicolumn{3}{c}{Ind. ARI} && \multicolumn{2}{c}{Avg. ARI} \\
\cline{3-4}\cline{6-7}\cline{9-11}\cline{13-14}
  && $>$ & $ = $ && AD-GMCM & PEM  && $>$ & $=$ & $<$  && AD-GMCM & PEM \\  
\midrule
1     && 22 & 8  && 147.4 & 146.4   && 10 & 9   & 11         && 0.552 & 0.554        \\ 
2     &&  19  & 11  && 146.9 & 145.9   &&  11  & 12 & 7      &&  0.570  & 0.568     \\
3     && 27 & 3  && 271.5 & 271.0   && 5 & 23   & 2         && 0.922 & 0.920        \\ 
4     &&  17  & 13 && 214.2 & 213.5    &&  11  & 15 & 4      &&  0.839  & 0.834     \\

5    &&  12 &  18 && 136.7 & 136.2  &&  5 & 19   & 6   &&  0.461 &    0.467  \\
6    &&  12 &  18 && 136.7 & 136.3  &&  5 & 20   & 5   &&  0.432 &    0.432  \\

7    &&  12 &  18 && 203.4 & 203.0  &&  3 & 25   & 2   &&  0.737 &    0.738  \\
8    &&  16 &  14 && 203.4 & 202.4  &&  10 & 17   & 3   &&  0.753 &    0.740  \\ \hline
Total && 137 & 103 & & & & &  60 & 140 & 40 & & &\\ 

\bottomrule
\end{tabular}
\caption{
Log-likelihood (LL) and ARI for AD-GMCM vs PEM over 8 different simulation settings for non-Gaussian mixtures. Indivdual (Ind.) results show the number of datasets in which AD-GMCM obtains higher, equal or lower values compared to PEM. Average (Avg.) results are over 30 datasets in each setting.
Details of simulations are given in tables \ref{tab:non_gmcm_sim_details_1} and \ref{tab:non_gmcm_sim_details_2}  }
\label{tab:non_gmcm_sim_results}
\end{table}

Table \ref{tab:non_gmcm_sim_results}
shows the number of times  AD-GMCM obtains higher, equal and lower log-likelihood and ARI, compared to PEM for each setting.
The average likelihood and ARI values, in each setting, are also shown. 
We observe that when the correlation is strong (settings 2,4,6,8), AD-GMCM performs better than PEM, in terms of average ARI.
The average ARI is higher than that of PEM in 3 out of 4 settings and is equal in the other setting.
When the correlation is weak (settings 1,3,5,7), the average ARI of PEM is higher in 3 out of 4 settings, despite having a lower average likelihood.
The number of datasets, in these settings, in which AD-GMCM obtains a higher ARI is 23, while PEM obtains a higher ARI in 21 datasets.
Because the GMCM model is misspecified for this set of simulations (i.e. the process from which the data is sampled and the model being fit are not the same), better likelihood does not necessarily imply better a clustering performance of the fitted model.
Overall, the number of datasets in which AD-GMCM obtains a higher ARI is 60, while PEM obtains a higher ARI in 40 datasets.

\subsection{Reproducibility Analysis}

Since reproducibility analysis is a special case of GMCM, AD-GMCM is expected to obtain better parameter estimates than PEM based on our simulation studies discussed above. 
\citep{bilgrau2016gmcm} note that in the case of reproducibility analysis PEM gets stuck in local maxima if the initializations are not close to the actual parameters.
So, we investigate this particular setting to compare PEM and HD-GMCM.

We simulate 30 2-dimensional datasets using the parameter values $(\alpha_1, \mu, \sigma, \rho) = (0.25,0.5,2,0.25)$. 
We evaluated PEM and HD-GMCM for 3 different initialization values -- set I:  $(0.32,0.5,1,0.25)$, set II: $(0.25,0.5,1,0.25)$ and set III: $(0.25,0.5,2,0.25)$ which are increasingly closer to the true parameters. 
Table \ref{tab:clus_accu_rep} shows
the number of times  AD-GMCM obtains higher, equal and lower log-likelihood and ARI, compared to PEM for each initialization setting.
The average likelihood and ARI values, for each setting, are also shown. 

From Table \ref{tab:clus_accu_rep}, we observe that in all the different initializations, AD-GMCM outperforms PEM in both likelihood and ARI. 
Further, it can be seen that the clustering performance of PEM is close, but inferior, to that of AD-GMCM only when the initialization values are close to the true parameters.  
These simulations strongly suggest that 
AD-GMCM escapes the local maxima in which PEM gets stuck in when parameter initializations are not close to that of the true parameters. 

\begin{table*}[ht!]
\centering
\begin{tabular}{c c cc c ccc c cc c cc}
\toprule
 Setting && \multicolumn{2}{c}{LL} && \multicolumn{2}{c}{Avg. LL} && \multicolumn{3}{c}{Ind. ARI} && \multicolumn{2}{c}{Avg. ARI} \\
 \cline{3-4}\cline{6-7}\cline{9-11}\cline{13-14}
  && $>$ & $\leq$ && AD-GMCM & PEM  && $>$ & $=$ & $<$  && AD-GMCM & PEM \\  
\midrule
I     && 29 & 1  && 5.815 & 4.325     && 18 & 3   & 9       && 0.037 & - 0.040        \\ 
II     &&  28  & 2   && 6.084  & 4.157  && 18   & 2 & 10      &&  0.039  & - 0.0231     \\ 
III   &&  21 &  9  &&  6.353 & 4.809  &&  15 & 6   & 9  &&  0.086 &    0.061  \\
\hline
Total && 78 & 12 & & & & &  51 & 11 & 28 & & &\\ 
\bottomrule
\end{tabular}
\caption{
Log-likelihood (LL) and ARI for AD-GMCM vs PEM over 3 different initialization settings for Reproducibility Analysis. 
Settings I--III have initializations increasingly closer to the true GMCM parameters.
Individual (Ind.) results show the number of datasets in which AD-GMCM obtains higher, equal or lower values compared to PEM. Average (Avg.) results are over 30 datasets in each setting. 
}
\label{tab:clus_accu_rep}

\end{table*}

\section{Real Datasets}
\label{sec:real}

\subsection{Cleveland Heart Disease Dataset}

\begin{figure}[!htbp]
        
\centering    
    \begin{subfigure}[b]{0.65\textwidth}
    \centering
\includegraphics[width= \textwidth]{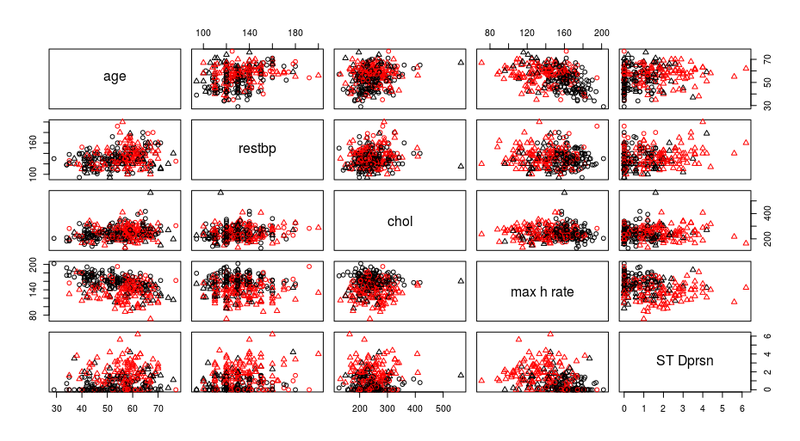}
        \caption{Actual Observations}
        
    \end{subfigure}%
  
 \begin{subfigure}[b]{0.65\textwidth}
 \centering
\includegraphics[width=\textwidth]{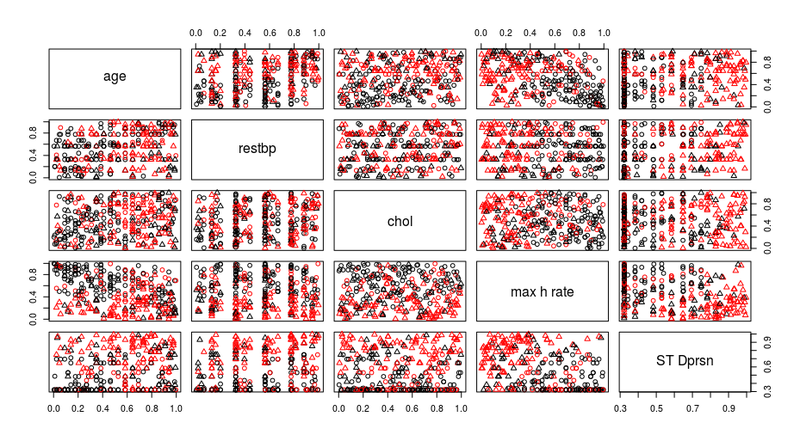} 
        \caption{Scaled Rank values}
\end{subfigure}%
    
 \begin{subfigure}[b]{0.65\textwidth}
 \centering
\includegraphics[width=\textwidth]{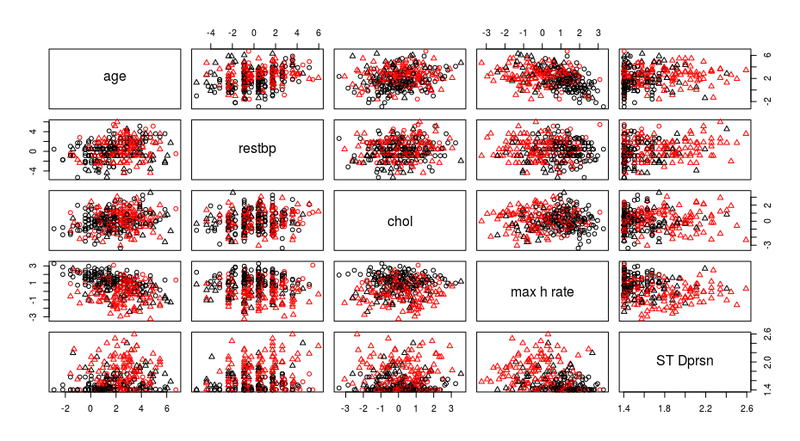} 
        \caption{Latent observation values}
        
    \end{subfigure}
   
    \caption{Pairwise scatterplots of features of patients \textit{with} (in red) and \textit{without} (in black) heart disease: input data (a), scaled rank values (b) and latent observations (c). There is a clearer cluster separation in Latent values and Scaled Rank values. The final clustering solution from GMCM clearly identifies patients with different maximum heart rate and ST Depression values. }
     \label{fig:scatter_cleveland}
    \end{figure}

The Cleveland Heart Disease dataset  measures risk factors for heart disease in 297 patients \citep{aha1988instance}. 
The task is to predict heart disease based on the collected attributes.
We use 5 numerical features -- age, resting blood pressure, serum cholestrol, maximum heart rate, ST depression levels.
Figure \ref{fig:scatter_cleveland} (a) shows the pairwise scatter plots of the features for patients \textit{with} (in red) and \textit{without} (in black) heart disease.
It can be seen that patients with heart disease have a lower maximum heart rate and higher ST depression values compared to patients without heart disease.  

On fitting GMCM with AD-GMCM, we observe that the model is primarily discriminating on the variables maximum heart rate and ST depression values. We observe that transforming observations to the latent space leads to relatively well separated Gaussian components for these two variables, as shown in  
the scatter plots in figure \ref{fig:scatter_cleveland}.

\begin{figure}[!h]
\centering
\begin{minipage}{\textwidth}
\centering
    \begin{subfigure}[b]{0.45\textwidth}
\includegraphics[width= \textwidth]{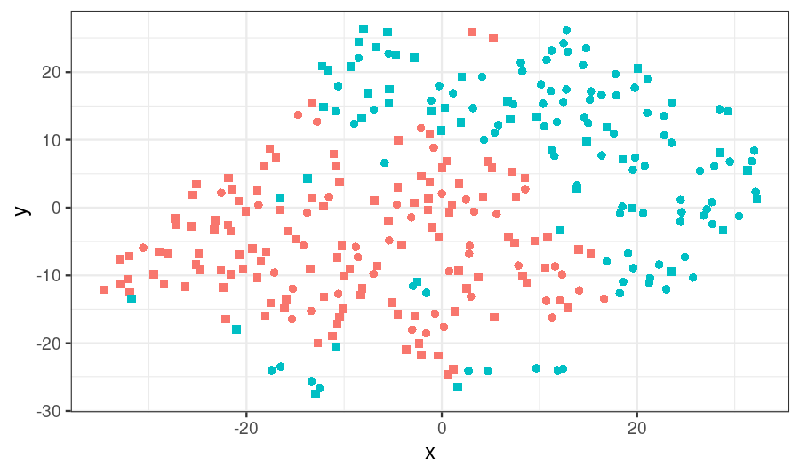}
        \caption{Actual observations}
        
    \end{subfigure}%
 \begin{subfigure}[b]{0.45\textwidth}
\includegraphics[width=\textwidth]{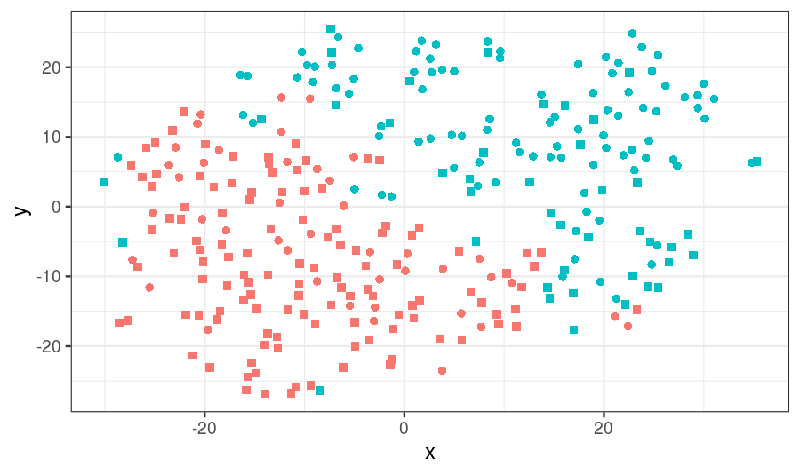} 
        \caption{Latent values}
        
    \end{subfigure}
 \begin{subfigure}[b]{0.6\textwidth}
\includegraphics[width=\textwidth]{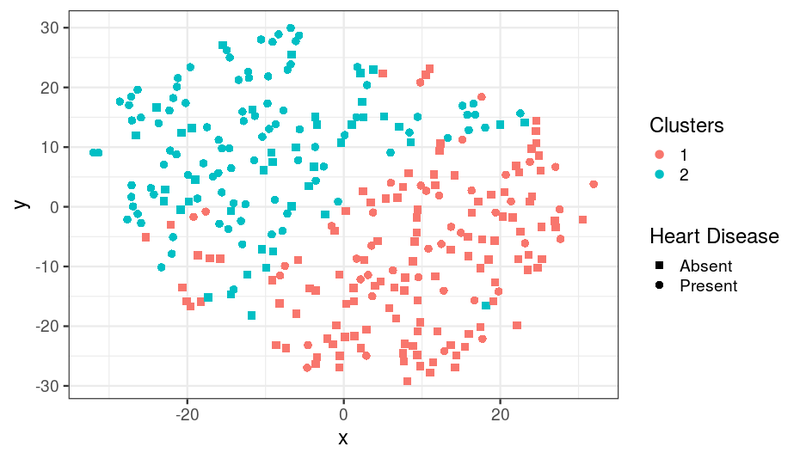} 
        \caption{Scaled Rank values}
            \end{subfigure}
\caption{tSNE plots of actual observations (a), latent observations (b) and scaled rank values (c). Transforming the actual observations to the latent space of GMCM leads to better cluster separation.}
\label{fig:tsne_cleveland}
\end{minipage}
\end{figure}

We also visualize 2-dimensional representations of all 5 features using tSNE \citep{maaten2008visualizing} of the clustering solution of AD-GMCM in figure \ref{fig:tsne_cleveland}. We observe that when the observations are transformed to the latent space, the cluster separation is more clearly visible. This highlights the advantage of GMCM for clustering.

Unsupervised classification of the patients using these features are done using AD-GMCM and PEM.
In addition, we also compare the performance with other widely used clustering algorithms such as GMM-EM, K-Means \citep{lloyd1982least}, Spectral clustering \citep{friedman2001elements} and Mixture of Factor Analyzers \citep{Ghah:Hilt:1997}.
All the clustering algorithms are evaluated using Adjusted Rand Index (ARI), where the ground truth (binary cluster labels) is based on presence/absence of heart disease.
The clustering performance of all the algorithms is shown in Table \ref{tab:cleveland_ari}. 
AD-GMCM has the highest ARI.
The PEM solution has a copula log likelihood of 80.59 whereas AD-GMCM solution has a much higher value of 94.24 for the same initialization, indicating  that AD-GMCM is better than PEM at escaping local optima.
\begin{table}[!h]
    \centering
    \begin{tabular}{|c|c|c|c|}
       \hline Algorithm & ARI & Algorithm & ARI \\
\hline AD-GMCM & $\mathbf{0.18}$ & K-Means & 0.168 \\
\hline PEM & 0.071 & Spectral & 0.019 \\
\hline GMM & 0.044 & MFA & 0.073 \\
\hline
    \end{tabular}
    \caption{ARI obtained by clustering algorithms on the Cleveland Heart Disease Dataset}
    \label{tab:cleveland_ari}
\end{table}

\subsection{U133VsExon Dataset}
The u133VsExon dataset \citep{bilgrau2016gmcm}
consists of p-values for tests conducted using two different microarray technologies.
The data matrix has 19,577 rows corresponding to the genes tested and 2 columns.
Two different microarray technologies -- Affymetrix GeneChip HG-U133 2.0 plus (U133) and GeneChip Human Exon 1.0 ST (Exon) -- are used to evaluate the same hypothesis, viz., the expression levels of the genes differ across two subtypes of lymphocyte cells. 
These two subtypes are known to play different roles in immune response to infections, which in turn has implications in blood cancer studies.
The aim is to use the data matrix to identify two groups of genes that are \textit{reproducibly} differentially expressed.

We fit the special case of GMCM for Reproducibility Analysis using both PEM and AD-GMCM. 
We tested for 5 different initializations and recorded the number of iterations taken to converge by PEM and AD-GMCM. The results are presented in Table \ref{tab:real_data_rep}.
\begin{table*}[h!]
\centering
\begin{tabular}{c c c c cc c cc}
\toprule
 S.No. && Initialization && \multicolumn{2}{c}{PEM} && \multicolumn{2}{c}{AD-GMCM}  \\
 \cline{5-6}\cline{8-9}
&&  && LL & Iterations   && LL & Iterations  \\  
\midrule
1 && (0.32,0.5,1,0.25)     && 4256.05 & 750     && 4317.12 & 750      \\ 
2 && (0.25,0.5,1,0.25)     &&  2858.01  & 21     &&  4296.68   & 750    \\ 
3 && (0.25,0.5,2,0.25)    &&  4254.95 &  750   &&  4316.68 & 750   \\
4 && (0.73,0.75,1.25,0.15)    &&  \textbf{4906.33} &  90   && \textbf{ 4910.69} & 750   \\
5 && (0.25,0.5,2,0.25)    &&  4905.65 &  125   &&  4909.50 & 750   \\
\bottomrule
\end{tabular}
\caption{Reproducibility Analysis on u133vsExon dataset: Best Log-likelihood (LL) and number of iterations taken by AD-GMCM and PEM at 5 different initializations.}

\label{tab:real_data_rep}
\end{table*}

We observe that AD-GMCM reaches a better log-likelihood for all the different initializations. 
The best estimated parameters by AD-GMCM are $(0.711,-1.801,1.298,0.767)$ which gives the highest log-likelihood of 4910.69.
The idr and adjusted idr values are computed with the get.IDR function of GMCM package \citep{bilgrau2016gmcm}. A total of 3505 genes (17.9\%) were found to have an adjusted IDR value below 0.05 and deemed reproducible. 
If MAP estimate (i.e., local idr $< 0.5$) is used, we found that 4504 genes (23\%) are reproducible. This is consistent with the estimate of mixture weight obtained (0.711) for the component 1 for which the null hypothesis is accepted. 

\begin{figure}[!h]
    \centering
    \begin{minipage}{.65\textwidth}
        \centering
    \begin{subfigure}[b]{\textwidth}
\includegraphics[width= \textwidth]{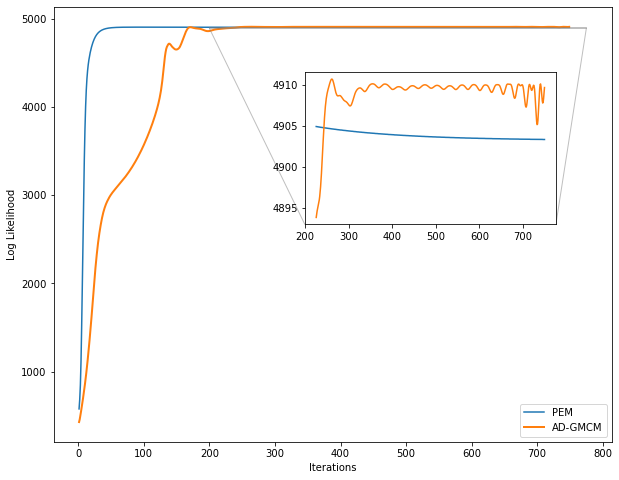}
 
    \end{subfigure}%
    
    \caption{Convergence of PEM and AD-GMCM in the u133VsExon dataset }
    \label{fig:U133_convergence}
    \end{minipage}
    \end{figure}

Figure \ref{fig:U133_convergence} shows the convergence plot for the fourth
initialization for both PEM and AD-GMCM.
Both PEM and AD-GMCM are first order methods, but PEM reaches a local maxima much faster compared to AD-GMCM. This is because PEM, being a variant of EM, has super-linear convergence \citep{jordanEMasGD}. However, it does not lead to a better log-likelihood as compared to AD-GMCM.

\begin{figure}[!h]
    \centering
    \begin{minipage}{.995\textwidth}
        \centering
    \begin{subfigure}[b]{0.5\textwidth}
\includegraphics[width= \textwidth]{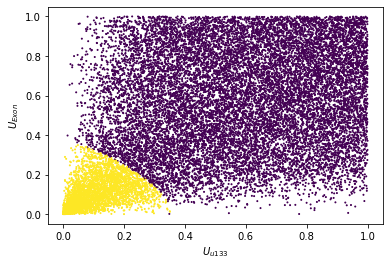}
        \caption{Scaled Rank Values}    
    \end{subfigure}%
    \begin{subfigure}[b]{0.5\textwidth}
\includegraphics[width=\textwidth]{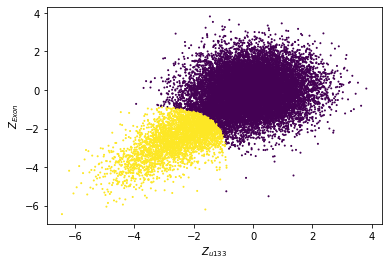}
        \caption{Latent Pseudo Observations}  
    \end{subfigure}
    \caption{Reproducibility analysis of U133vsExon dataset. Reproducible points are in yellow.}
    \label{fig:U133_scatter}
    \end{minipage}
    \end{figure}
\begin{figure}[!h]
    \centering
    \begin{minipage}{.995\textwidth}
        \centering
    \begin{subfigure}[b]{0.5\textwidth}
\includegraphics[width= \textwidth]{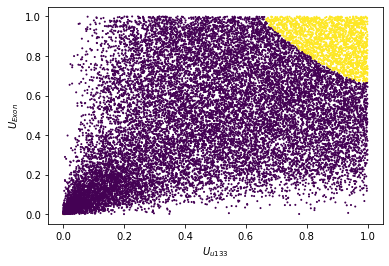}
        \caption{Scaled Rank Values}    
    \end{subfigure}%
    \begin{subfigure}[b]{0.5\textwidth}
\includegraphics[width=\textwidth]{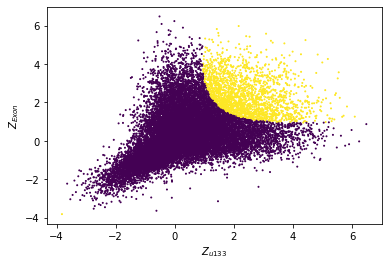}
        \caption{Latent Pseudo Observations}  
    \end{subfigure}
    \caption{Reproducibility analysis of U133vsExon dataset, using estimated parameters from \cite{bilgrau2016gmcm}. Reproducible points are in yellow.}
    \label{fig:U133_scatter_bilgrau}
    \end{minipage}
    \end{figure}

Figure \ref{fig:U133_scatter} shows the scaled rank values and latent observations inferred by AD-GMCM for the fourth initialization that obtains the highest likelihood of 4910.69.
We plot the observations corresponding to the reproducible genes, identified using the adjusted IDR, in figure \ref{fig:U133_scatter} in yellow. 
For a comparison, figure \ref{fig:U133_scatter_bilgrau} shows the plot for the solution reported by \cite{bilgrau2016gmcm} for the same dataset.
We compute the likelihood of the GMCM model using the estimated parameters reported by \cite{bilgrau2016gmcm}: $(0.71,1.83,1.31,0.76)$, to find a much
lower log-likelihood of -4997.84.
We identify the reproducible points by setting the same threshold of 0.05 on the adjusted IDR, and plot the corresponding observations in figure \ref{fig:U133_scatter_bilgrau} in yellow. Note that the reproducible points identified by \citep{bilgrau2016gmcm} are predominately those with high p-values in both the experiments (upper right corner in figure \ref{fig:U133_scatter_bilgrau}). 
In figure \ref{fig:U133_scatter}, the solution from AD-GMCM, with higher likelihood, clearly shows a weakly correlated component 1 (in purple) and a strongly correlated component 2 (in yellow, lower left corner) comprising reproducible genes with low p-values in both experiments.

\section{Discussion and Conclusion}
\label{sec:conclusion}

In this paper we present a new algorithm 
to infer the parameters of the Gaussian Mixture Copula Model (GMCM). 
Due to the challenges of maximizing the exact GMCM likelihood, the best previous inference algorithms maximize the pseudo-likelihood to obtain parameter estimates, through a Pseudo Expectation Maximization (PEM) approach.
Our approach, AD-GMCM, is based on Automatic Differentiation that provides accurate computation of the gradients and obviates the need to calculate them in closed form.
As a result, AD-GMCM can be implemented easily and used to maximize the exact GMCM likelihood.

We conducted extensive experiments on simulated datasets to compare AD-GMCM and PEM.
Our experiments show that AD-GMCM obtains more accurate parameter estimates than PEM.
As a result, AD-GMCM obtains improved clustering of data generated by GMCM.
AD-GMCM also obtains better likelihood fit and clustering solutions for other non-Gaussian mixtures tested.
In Reproducibility Analysis, that uses a special case of GMCM, a known limitation of PEM is its inability to escape local maxima when initializations are not close to the true parameters. 
We find that AD-GMCM does not have this limitation and obtains good results even when initializations are far from the true GMCM parameters.
Our experiments on two real datasets demonstrate the efficacy of AD-GMCM in clustering and Reproducibility Analysis.

We theoretically analyse conditions for monotonic increase in copula likelihood and derive new conditions for the inverse CDF values that are estimated in each iteration of both PEM and AD-GMCM.
These conditions are less restrictive than those derived previously by \citep{compstat,kasa2020gaussian}. 
However, we also find empirically that a simpler alternative, in a gradient-based approach, is to set a small learning rate.
We illustrate the problem of unidentifiability of component means and weights of GMCM. 
Previous approaches to address this problem can easily be incorporated into AD-GMCM.

Thus, we find that the AD-based approach to estimating GMCM parameters addresses the challenge of maximizing the exact GMCM likelihood with relative ease of implementation and outperforms previous approaches in parameter estimation, clustering and reproducibility analysis.
An implementation of our approach is available for public use\footnote{Undisclosed to preserve anonymity during double blind review}
Our study illustrates the benefits of Automatic Differentiation, that is well known but rarely used, for parameter inference in statistical models.
We believe that this study shall provide the impetus to explore the use of AD in parameter estimation of other copula-based models as well.

We analyze two cases that lead to degenerate solutions during Maximum Likelihood estimation of Gaussian Mixture Models, that in turn can lead to singular component covariance matrices and clustering solutions with very few closely located data points.
We find, surprisingly, that degenerate solutions due to component covariance becoming arbitrarily small -- a well-known case of degeneracy for GMM -- does not occur in GMCM. To our knowledge, this has not been reported in previous literature.

Identifying the possible causes of degeneracy in the likelihood has practical significance in detecting spurious solutions, which are local maximizers of the likelihood function but lack real-life interpretability and hence do not provide a good clustering of the data \citep{McLa:Peel:fini:2000}.
Constraining the parameter space has been the primary approach to avoid spurious solutions in GMM \citep{hathaway1986constrained,ingrassia2004likelihood,Ingr:Rocc:2007,Chen2009,ingrassia2011degeneracy}. With respect to GMCM, the relative occurrence of spurious solutions for different inference techniques, initialization strategies, parameterizations, and with increasing dimensionality can be investigated in the future. 
Accurate computation of inverse CDF of GMM through function approximators and
constructing algorithms to obtain the least restrictive modifications that ensure a monotonic increase in the GMCM likelihood can be explored to further improve GMCM parameter inference.

\bibliography{biblio}

\newpage
\appendix
\section{Symbols and Notation}
\label{sec:symbols}

	\begin{table}[!h]
	\centering
\begin{tabular}{cp{12cm} } 

 \hline
 Symbol &  Meaning  \\ \hline
 $n$ & Total number of $iid$ datapoints considered \\
 $p$ & Total number of dimensions or features of the data \\
 $p_e$ & Total number of free parameters in the model \\
  $K$ & total of number of components in the model \\
 $\phi$ & density of multivariate Gaussian distribution \\
 $f$ & overall multivariate probability density from which data is sampled\\
 $F$ & overall multivariate CDF \\
 $f_j$ & marginal probability density along the $j$-th component \\
  $\bF_j$ & marginal CDF along the $j$-th dimension \\
 $c$ & copula density \\
 $\mathcal{C}$ & copula likelihood corresponding to the density $c$ \\
 $\mathcal{L}$ & Log likelihood of the copula density $c$ \\
  $\mathcal{L}_p$ & Pseudo Log likelihood corresponding to the copula density $c$ \\
 $\btheta$ & all parameters of the model\\
  $f_{\text{GMM}}$ & density of the latent GMM \\
  $\Psi_{j}$ & marginal CDF of latent GMM along $j$-th dimension \\
   $\psi_{j}$ & marginal density of latent GMM along $j$-th dimension \\
 $\bTheta$ & set of all possible parameters i.e. the entire parameter space \\
$k$ & indices over the total number of components $K$  \\
$j$ & indices over the total number of dimensions $p$  \\
$i$ & indices over the total number of datapoints $n$  \\
 $\bX_i^j$ & $j$-th dimension of $i$-th random datapoint \\
 $\bU_i^j$ & Scaled rank along $j$-th dimension of $i$-th random datapoint \\
 $z_k^i$ & latent random variable corresponding to the cluster membership of $i$-th observation to the $j$-th cluster \\
 $\pi_k$ & mixing proportions or weight of the $k$-th component \\
 $\bmu_k, \mu_{k}^{j} $ & mean of $k$-th component of a GMM and its value of along $j$-th dimension respectively \\
 $\bSigma_k, \Sigma_{kij}$ & Covariance matrix of $k$-th component of a GMM and its $ij$-th element respectively \\
  $\bV_k$ & square-root of $k$-th covariance matrix $\bSigma_k$\\
  $\hat{a}^{(t)}$ &  Estimate of parameter $a$ at the end of iteration $t$ \\ 
  $\lambda$ & factor controlling the cluster separation in simulations \\ 
  $\mathbb{I}$ & Identity matrix \\ 
$\epsilon$ & learning rate in the vanilla gradient descent step \\
$\gamma$ & convergence threshold for stopping criterion \\
$|.|$ & absolute value \\
 \hline
 
\end{tabular}
\captionof{table}{Symbols used in the paper}
\label{table:symbols}
\end{table}

\section{Automatic Differentiation}
\label{app:AD}

The examples in this section are from \cite{baydin2018automatic}.
In symbolic differentiation, we first evaluate the complete expression and then differentiate it using rules of differential calculus. 
First note that a naive computation may repeatedly evaluate the same expression multiple times, e.g., consider the rules:
\begin{align}
\frac{d}{dx} \left(A(x) + B(x)\right) \leadsto \frac{d}{dx} A(x) + \frac{d}{dx} B(x) \\
\frac{d}{dx} \left(A(x)\,B(x)\right) \leadsto \left(\frac{d}{dx} A(x)\right) B(x) + A(x) \left(\frac{d}{dx} B(x)\right)
    \label{EquationMultiplicationRule}
\end{align}

Let $H(x)=A(x)B(x)$. 
Note that $H(x)$ and $\frac{d}{dx}H(x)$ have in  common: $A(x)$ and $B(x)$, 
and on the right hand side, $A(x)$ and $\frac{d}{dx}A(x)$ appear separately. 
In symbolic differentiation we plug the derivative of $A(x)$ and thus 
have nested duplications of any computation that appears in common between $A(x)$ and $\frac{d}{dx}B(x)$. 
In this manner symbolic differentiation can produce exponentially large symbolic expressions which take correspondingly long to evaluate. This problem is called \textbf{expression swell}.
To illustrate the problem, consider the following  iterations of the logistic map $l_{n+1}=4l_n (1-l_n)$, $l_1=x$ and the corresponding derivatives of $l_n$ with respect to $x$. Table \ref{TableExpressionSwell} clearly shows that the number of repetitive evaluations increase with $n$.

\begin{table}
  \centering
  \captionof{table}{Iterations of the logistic map $l_{n+1}=4l_n (1-l_n)$, $l_1=x$ and the corresponding derivatives of $l_n$ with respect to $x$, illustrating expression swell 
  (from \cite{baydin2018automatic}) }
  \label{TableExpressionSwell}
  \renewcommand{\arraystretch}{1}
  
  {\small
  \begin{tabularx}{\columnwidth}{@{}lXp{3cm}XX@{}}
    \toprule
    $n$ & $l_n$ & $\frac{d}{dx}l_n$ & $\frac{d}{dx}l_n$ (Simplified form)\\
    \addlinespace
    \midrule
    1 & $x$ & $1$ & $1$\\
    2 & $4x(1 - x)$ & $4(1 - x) -4x$ & $4 - 8x$\\
    3 & $16x(1 - x)(1 - 2 x)^2$ & $16(1 - x)(1 - 2 x)^2 - 16x(1 - 2 x)^2 - 64x(1 - x)(1 - 2 x)$ & $16 (1 - 10 x + 24 x^2 - 16 x^3)$\\
    4 & $64x(1 - x)(1 - 2 x)^2$ $(1 - 8 x + 8 x^2)^2$ & $128x(1 - x)(-8 + 16 x)(1 - 2 x)^2 (1 - 8 x + 8 x^2) + 64 (1 - x)(1 - 2 x)^2  (1 - 8 x + 8 x^2)^2 - 64x(1 - 2 x)^2 (1 - 8 x + 8 x^2)^2 - 256x(1 - x)(1 - 2 x)(1 - 8 x + 8 x^2)^2$ & $64 (1 - 42 x + 504 x^2 - 2640 x^3 + 7040 x^4 - 9984 x^5 + 7168 x^6 - 2048 x^7)$\\
    \bottomrule
  \end{tabularx}}
\end{table}

If the symbolic form is not required and only numerical evaluation of derivatives is required, computations can be simplified by storing the values of intermediate sub-expressions.
Further efficiency gains in computation can be achieved by interleaving differentiation and simplification steps.
The derivative of $l_{n+1} = 4 l_{n} (1- l_{n})$ can be found using the chain rule $\frac{dl_{n+1}}{dl_n}\frac{dl_{n}}{dl_{n-1}}\dots \frac{dl_{1}}{dl_x}$ which simplifies to $4(1-l_{n} - l_{n})4(1-l_{n-1} - l_{n-1})\dots4(1-x - x)$. 
Note that evaluation in AD is computationally linear in $n$ (because we add only one $(1-l_n - l_n)$ for each increase by 1).
This linear time complexity is achieved due to `carry-over' of the derivatives at each step, rather than evaluating the derivative at the end and substituting the value of x. 

\newpage
The Python code below shows the simplicity of the implementation for this problem.

\begin{center}
\begin{verbatim}
from autograd import grad
def my_func(x,n):
    p = x
    y = x * (1 - x)
    for i in range(n):
        y = y*(1 - y)
        
    return y
grad_func = grad(my_func)
grad_func(0.5,4)
\end{verbatim}
\end{center}

Consider the following recursive expressions: 
$l_0 = \frac{1}{1+e^x}$, $l_1 = \frac{1}{1+e^{l_0}} $, ....., $l_{n} = \frac{1}{1+e^{l_{n-1}}} $
We evaluate the derivative of $l_{n}$ with respect to $x$ and compare the runtime in Mathematica (SD) vs PyTorch (AD) for various values of $n$. As $n$ increases, it is expected that runtime also increases. It can be seen from the results in Table \ref{Runtime}  that runtime increases linearly for AD (using PyTorch) whereas it increases exponentially for SD (using Mathematica).

\begin{center}
\captionof{table}{Average runtime (over 1000 runs) }
\label{Runtime}
\begin{tabular}{ ccc } 

 \hline
 n & \text{AD}  & \text{SD} \\ \hline
1 & 0.00013 & 0.00000 \\ 
5 & 0.00030 & 0.00005 \\ 
10 & 0.00051 & 0.00023 \\
50 & 0.00293 & 0.00437 \\
100 & 0.00433 & 0.15625 \\
200 & 0.00917 & 1.45364 \\
 \hline
\end{tabular}
\end{center}

\section{Proof of Conditions for Monotonicity}\label{sec:proofmono}

\begin{proof}
Let $z_{k}^{i}$ be the latent variable which indicates the cluster membership of the $i$-th observation to the $k$-th component. 

The complete data $(X,Z)$ likelihood is given by

\begin{align}
   \mathcal{C}_c(\by,\bz) = \Pi_{n=1}^{N} \Pi_{k=1}^{K}  \frac{\left( \pi_k\phi(\by_i|\bmu_k,\bSigma_k) \right)^{z_k^{i}} }{\Pi_{j=1}^{p} \sum_k \left( \pi_k\psi_j(y_i^j|\mu_k^j,\Sigma_{kjj})\right)^{z_k^i}}
   \label{eqn:proof_copula_complete}
\end{align}

Taking logarithm on both sides of Eq. \ref{eqn:proof_copula_complete}, we can see the complete copula log likelihood is composed of the pseudo complete log likelihood and sum of marginal log likelihoods. We try to give the conditions for which the pseudo complete log likelihood increases and the marginal log likelihoods decrease.

 Even though \cite{kasa2020gaussian} give a condition under which the complete pseudo log likelihood $\mathcal{L}_p (= \ln (f) )$ increases with every iteration, we find that it is possible to construct a much less restrictive set of conditions.

\subsection{Pseudo Complete Log likelihood}

\begin{align*}
\log \mathcal{L}_{p} & (\btheta^{(t+1)} | \by^{(t+1)}, \bz^{(t)}) - \log  \mathcal{L}_{p} (\btheta^{(t+1)} | \by^{(t)}, \bz^{(t)})     = \\ &  -0.5\sum_{i=1}^{N}\sum_{k=1}^{K} \biggl(  {{z_k^i}^{(t)}} \biggl( {(\textbf{y}_{i}^{(t+1)}  - \vecmu_{k}^{(t+1)} )}^{T} \\ &{\bSigma_{k}^{(t+1)}}^{-1}   (\textbf{y}_{i}^{(t+1)}  - \vecmu_{k}^{(t+1)} )                                \biggr)   \\ &  - {z_k^{i^{(t)}} } \left(    {(\textbf{y}_{i}^{(t)}  - \vecmu_{k}^{(t+1)} )}^{T} {\bSigma_{k}^{(t+1)}}^{-1}   (\textbf{y}_{i}^{(t)}  - \vecmu_{k}^{(t+1)} )                                \right)          \biggr)   
\end{align*}
First, note that if $A$ is $1${$\times$}$1$, then $tr(A) = A$. Also, note that $tr(ABC) = tr(BAC)$. Using these facts, we can simplify the above expression to

\begin{align*}
 & -0.5  \sum_{i=1}^{N}\sum_{k=1}^{K} \biggl(  {z_k^{i^{(t)}}} \biggl( {(\textbf{y}_{i}^{(t+1)}  - \vecmu_{k}^{(t+1)} )}^{T} \\&{\bSigma_{k}^{(t+1)}}^{-1}   (\textbf{y}_{i}^{(t+1)}  - \vecmu_{k}^{(t+1)} )                                \biggr)  \\  - &{z_k^{i^{(t)}} } \biggl(    {(\textbf{y}_{i}^{(t)}  - \vecmu_{k}^{(t+1)} )}^{T} {\bSigma_{k}^{(t+1)}}^{-1}   (\textbf{y}_{i}^{(t)}  - \vecmu_{k}^{(t+1)} )                                \biggr)          \biggr)  \\   = & -0.5  \sum_{i=1}^{N}\sum_{k=1}^{K}  \biggl( {z_k^{i^{(t)}} } tr\biggl( {\bSigma_{k}^{(t+1)}}^{-1}   {(\textbf{y}_{i}^{(t+1)}  - \vecmu_{k}^{(t+1)} )}^{T} \\&{(\textbf{y}_{i}^{(t+1)}  - \vecmu_{k}^{(t+1)} )}   \biggr) \\ & -  {z_k^{i^{(t)}} } tr\left( {\bSigma_{k}^{(t+1)}}^{-1}   {(\textbf{y}_{i}^{(t)}  - \vecmu_{k}^{(t+1)} )}^{T} {(\textbf{y}_{i}^{(t)}  - \vecmu_{k}^{(t+1)} )}   \right)\biggr)  \\  & = -0.5  \sum_{i=1}^{N}\sum_{k=1}^{K}  \biggl( {z_k^{i^{(t)}}} tr\left({\bSigma_{k}^{(t+1)}}^{-1}\right) \\ &  \underbrace{ ( -2 {(\vecmu_{k}^{(t+1)})}^{T} ( \textbf{y}_{i}^{(t+1)} - \textbf{y}_{i}^{(t)} ) + ( {\textbf{y}_{i}^{(t+1)}}^{T} \textbf{y}_{i}^{(t+1)} - {\textbf{y}_{i}^{(t)}}^{T} \textbf{y}_{i}^{(t)} ) )}_{\text{Say A}} \biggr)  
\end{align*}

Note that by construction, ${z_k^{i^{(t)}}}$ is always non-negative. Also, since ${\bSigma_{k}^{(t+1)}}^{-1}$ is positive definite by construction, so its trace is always positive. Therefore, for the likelihood to increase the term A has to be negative. The term A is composed of inner products which we individually set $\leq 0$

Taking the summation over $k$, we have the following $ij$ quadratic inequalities
\begin{align}
  k \biggl( {{y_i^j}^{t+1}}^2 - {{y_i^j}^{t}}^2 \biggr) - 2 \sum_k \mu_k^j \biggl( {y_i^j}^{t+1} - {y_i^j}^{t} \biggr) \leq 0 \qquad \qquad \forall {i,j}
\end{align}
which yields the following conditions

\begin{enumerate}
    \item If ${y_i^j}^{(t+1)} - {y_i^j}^{(t)} > 0$ then ${y_i^j}^{(t+1)} < \frac{2\sum_k \mu_k^j}{k} - {y_i^j}^{(t)} $
\item If ${y_i^j}^{(t+1)} - {y_i^j}^{(t)} < 0$ then ${y_i^j}^{(t+1)} > \frac{2\sum_k \mu_k^j}{k} - {y_i^j}^{(t)} $

\end{enumerate}

\subsection{Marginal log likelihood}

In order to ensure that marginal log likelihoods decrease with every iteration, we set 

\begin{align}
\left( \ln \left( 
e^{-\frac{ (y_i^{j^{(t+1)}} - \mu_k^j)^2}{2\Sigma_{kjj}^2}}
\right) - \ln \left( 
e^{-\frac{ (y_i^{j^{(t)}} - \mu_k^j)^2}{2\Sigma_{kjj}^2}}
\right) \right) \leq 0
\label{eqn:proof_copula_marginal}
\end{align}

From this we can see that a sufficient condition to ensure that Eq. \ref{eqn:proof_copula_marginal} holds is 
\begin{align}
    (y_i^{j^{(t+1)}} - \mu_k^j)^2 \ge (y_i^{j^{(t)}} - \mu_k^j)^2
\end{align}

Now, the overall log likelihood $\mathcal{L}(\btheta | \by)$ also increases monotonically because 

\begin{align*}
    \mathcal{L}(\btheta^{(t+1)} | \by^{(t+1)}) &  = \sum_{\bz} \mathcal{L}(\btheta^{(t+1)} | \by^{(t+1)}, \bz^{(t)})  \\ & \ge  \sum_{\bz} \mathcal{L}(\btheta^{(t+1)} | \by^{(t)}, \bz^{(t)})  \\ & = \mathcal{L}(\btheta^{(t+1)} | \by^{(t)})
\end{align*}

\end{proof}

\section{High Values of GMCM likelihood}
\label{sec:highLL}

Observe that the GMCM likelihood in Eq. \ref{eq:exactGMCM} is composed of pseudo likelihood and sum of marginal likelihoods along each dimension. Taking log on both sides, we have 
\begin{equation}
\mathcal{L} = \sum_{i=1}^n \underbrace{\log  \biggl( \sum_{k=1}^K \pi_k\phi(\vecy_i\mid\bmu_k,\bSigma_k)\biggr)}_{\mathcal{L}_p} - \sum_{i=1}^n \sum_{j=1}^p \log \biggl(  \sum_{k=1}^{K} \psi_j(y_{i}^j \mid\mu_k^j,\Sigma_{kjj} )) \biggr)
\end{equation}
Typically, we expect the sum of marginal distributions (the second term) to not have better fit than the multivariate GMM log-likelihood or the pseudo log-likelihood (the first term, $\mathcal{L}_p$) because the marginal distributions do not take into account the component correlation structures. While both these log likelihoods are individually negative, the overall GMCM log likelihood is usually positive because the marginal log likelihoods (large -ve values) are being subtracted from the pseudo log likelihood.

\end{document}